\documentclass[iop,revtex4]{emulateapj}
\usepackage{color}
\usepackage{amsmath}
\usepackage{xcolor}
\usepackage{mathrsfs}
\usepackage{lscape}
\usepackage{natbib}
\usepackage{gensymb}

\shortauthors{Bryan et al.}
\shorttitle{As the Worlds Turn}
\slugcomment{{\sc } October 12, 2020}

\begin{document}

\title{As the Worlds Turn:  Constraining Spin Evolution in the Planetary-Mass Regime}

\author{
Marta L. Bryan\altaffilmark{1}, Sivan Ginzburg\altaffilmark{1}, Eugene Chiang\altaffilmark{1}, Caroline Morley\altaffilmark{2}, Brendan P. Bowler\altaffilmark{2}, Jerry W. Xuan\altaffilmark{3}, Heather A. Knutson\altaffilmark{4}
}

\altaffiltext{1}{Department of Astronomy, 501 Campbell Hall, University of California Berkeley, Berkeley, CA 94720-3411, USA}

\altaffiltext{2}{Department of Astronomy, The University of Texas at Austin, Austin, TX 78712, USA}

\altaffiltext{3}{Cahill Center for Astronomy and Astrophysics, California Institute of Technology,
1200 East California Boulevard, MC 249-17, Pasadena, CA 91125, USA}

\altaffiltext{4}{Division of Geological and Planetary Sciences, California Institute of Technology, 1200 East California Boulevard, MC 150-21, Pasadena, CA 91125, USA}

\begin{abstract} 
 
To understand how planetary spin evolves and traces planet formation processes, we measure rotational line broadening in eight planetary-mass objects (PMOs) of various ages (1--800 Myr) using near-infrared high-resolution spectra from NIRSPEC/Keck.  Combining these with published rotation rates, we compile 27 PMO spin velocities, 16 of which derive from our NIRSPEC/Keck program.  Our data are consistent with spin velocities $v$ scaling with planetary radius $R$ as $v \propto 1/R$. We conclude that spin angular momentum is conserved as objects cool and contract over the sampled age range. The PMOs in our sample spin at rates that are approximately an order of magnitude below their break-up values, consistent with the hypothesis that they were spun down by magnetized circum-PMO disks (CPDs) during the formation era at ages $\lesssim$ a few Myr.  There is a factor of 4--5 variation in spin velocity that has yet to be understood theoretically. It also remains to be seen whether spin evolves on timescales $\gtrsim$ 1 Gyr for PMOs, as it does for stars and high-mass brown dwarfs emitting magnetized winds.

\end{abstract}

\section{Introduction}

Spin is an observable that informs our understanding of planet formation and evolution.  Whether planetary-mass objects (PMOs, defined here as having masses near or below the deuterium-burning limit) form bottom-up (via core accretion) or top-down (via gravitational instability), they
initially contain as much angular momentum as is carried by the gas they accrete. That angular momentum, if strictly conserved,
could force newly accreted PMOs to rotate at near break-up speeds, halting their contraction at sizes much larger than a Jupiter radius (e.g., \citealt{Ginzburg2020}). The observational fact that PMOs and gas giants are not so distended suggests their spin angular momenta were regulated or shed. The need to remove angular momentum to enable contraction is called out in some hydrodynamical
simulations 
(e.g., \citealt{Lambrechts2019};
cf. \citealt{Szulagyi2016}); 
the numerical evidence in
either direction is limited so far because simulations typically do not follow the contraction down to the centrifugal barrier,
and because inadequate resolution when the planet contracts significantly inside its Hill sphere can artificially
reduce the angular momentum
budget.

In \citet[][hereafter B18]{Bryan2018}, we found that eleven PMOs spin at speeds well below break-up,
at ages as young as $2 \pm 1$ Myr. This finding is consistent with objects having spun down early by  primordial, magnetized, circum-PMO disks (CPDs).  Here a CPD can refer either to a disk surrounding an isolated PMO, or a disk surrounding a PMO which itself is a companion to another star.  The idea that an object's rotation rate can be locked to the rotation rate of a magnetospherically truncated disk has been applied to neutron stars \citep[e.g.][]{Ghosh1979,Romanova2016}, T Tauri stars \citep[e.g.][]{Konigl1991,Ostriker1995}, Solar System gas giants \citep[e.g.][]{Takata1996}, and extrasolar PMOs \citep[e.g.][]{Batygin2018}.  \citet{Ginzburg2020} found that the magnetic braking timescale is shorter than the Kelvin-Helmholtz timescale over which young planets cool and contract; maintaining this inequality is crucial for enabling planets to shrink to their observed sizes, on the order of a Jupiter radius.

Here we measure eight new rotation speeds of PMOs. Adding these to the \citet{Bryan2018} sample and other published rotation rates yields a total of 27 PMO spin measurements, 14 of bound planetary-mass companions (PMCs) and 13 of free-floating low-mass ($<$ 20 M$_{\rm Jup}$) brown dwarfs. While the bound companions likely formed in a disk, the isolated low-mass brown dwarfs are thought to form via molecular cloud fragmentation. Both formation pathways lead to accretion disks (CPDs), either around a bound or an isolated PMO. Our new spins derive from rotational line broadening as measured from our ongoing NIRSPEC/Keck high-resolution spectroscopic survey. Our target PMOs are directly imaged objects that are relatively young (typically $\lesssim$ 100 Myr), massive ($\sim$$10 M_{\rm Jup}$),and far ($\gtrsim$ 50 AU,i.e., $\gtrsim$ 1 arcsec) from their host stars.  These characteristics better enable either spectroscopy of the PMO itself (which yields a $v\sin i$ measurement from rotational line broadening), or photometric monitoring of the planet (which yields a photometric rotation period) \citep[e.g.][]{Zhou2016,Zhou2019,Zhou2020}.
 
The rest of this paper is organized as follows.  In Section 2 we describe our NIRSPEC/Keck K-band observations.  Section 3 details how we extracted and reduced each spectrum, and how we measured rotational line broadening.  In Section 4 we consider the entire compilation of 27 PMO spin measurements and discuss how they constrain spin evolution in the planetary-mass regime.  We summarize in Section 5.

\section{NIRSPEC K-Band Observations}

All targets in our sample are empirically selected to have masses below 20 M$_{\rm Jup}$. Since objects in our study were discovered using direct imaging,  uncertainties on mass measurements are large. Because the bound companion population have masses that typically straddle the deuterium burning limit and whose 1$\sigma$ uncertainties approach 20 M$_{\rm Jup}$, this cutoff yields a mass distribution for the low-mass free-floating brown dwarfs that is consistent with the bound companion sample. We observed all of our targets in \textit{K} band (2.03 - 2.38 um) using the near-infrared high-resolution spectrograph NIRSPEC at the Keck II 10m telescope \citep{McLean1998}.  Since all observations were taken prior to the NIRSPEC upgrade in April 2018, the instrumental resolution of these spectra was $\sim$25,000.  For the ROXs12 system we carried out observations in adaptive optics (AO) mode using the 0.041 $\times$ 2.26 arcsec slit, in order to minimize blending of the light from the companion at a projected separation of 1$\farcs$8 from its the host star.  Since all other systems had companions that were sufficiently far away from their host stars ($>$8 arcsec) or were free-floating objects, we observed the remainder of our sample in natural seeing mode using the 0.432 $\times$ 24 arcsec slit.  This mode has a significantly greater ($\sim$10$\times$) throughput than AO mode.  For ROXs12 b and SR12 c, we placed the companion and the star in the slit to observe both simultaneously, which later facilitated wavelength solution calculations and telluric corrections for the spectra of the faint companions.  This strategy was not possible for 2M0249-0557c given that the companion-star separation (40 arcseconds) was larger than the slit length, so we observed each separately.  All data were taken using standard ABBA or AB nod sequences (depending on exposure length).  See Table \ref{tb:obs} for observation details.

\begin{deluxetable*}{cccccccccccc}
\tablecaption{ NIRSPEC \textit{K}-Band Observations \label{tb:obs}}
\tabletypesize{\footnotesize}
\tablehead{
  \colhead{System } & 
  \colhead{RA} &
  \colhead{Dec} &
  \colhead{Pri. SpT } & 
  \colhead{m$_{\rm K,\star}$ } & 
  \colhead{Pl. SpT } & 
  \colhead{m$_{\rm K,pl}$ } & 
  \colhead{Proj. Sep.} & 
  \colhead{UT Date} &
  \colhead{AO?} &
  \colhead{$\#$ Exp.} &
  \colhead{Tot. Exp.}\\
  \colhead{} & 
  \colhead{} & 
  \colhead{} &
  \colhead{} &
  \colhead{[mag]} & 
  \colhead{} & 
  \colhead{[mag]} & 
  \colhead{['']} &
  \colhead{} &
  \colhead{} &
  \colhead{} &
  \colhead{[min]}
}
\startdata
 ROXs 12 b& 16:26:28.03 & –25:26:47.7 & M0 & 9.1 & L0 & 14.1 & 1.8 & 2017 July 11 & Yes & 14  & 167 \\
 SR 12 c& 16:27:19.51 & -24:41:40.4 & K4, M2.5 & 8.4 & M9.0 & & 8.7 & 2017 July 2 & No & 24 & 127\\
 2M0249-0557 c& 02:49:56.39 & -05:57:35.3 & M6, M6 & 11.7, 11.9 & L2 & 14.8 & 40.0 & 2018 July 22 & No & 6 & 50 \\
 OPH 98& 16:27:44.23 & -23:58:52.1 & $\cdots$ &$\cdots$ & M9.75 & 15.0 &$\cdots$ & 2017 July 2 & No & 6 & 63 \\
 OPH 103& 16:28:10.45 & -24:24:20.1 &$\cdots$ &$\cdots$ & L0 & 15.0 & $\cdots$& 2017 July 28 & No & 6 & 73 \\
 2M2244+2043& 22:44:31.67 & +20:43:43.3 &$\cdots$ & $\cdots$& L6 & 14.0 &$\cdots$ & 2017 July 28 & No & 6 & 73 \\
 2M2013--2806& 20:13:51.53 & -28:06:02.0 &$\cdots$ &$\cdots$ & M9 &  12.9 & $\cdots$ & 2018 July 22 & No & 6 & 50 \\
 2M2208+2921& 22:08:13.63 & +29:21:21.5 &$\cdots$ & $\cdots$& L2 & 14.1 & $\cdots$ & 2018 July 22 & No & 6 & 60
\enddata
\tablecomments{See Table \ref{tb:res} for relevant references.}
\end{deluxetable*}

\section{Analysis}

\subsection{1D Spectrum Extraction}
Following the methodology detailed in \citet{Bryan2018}, we extract 1D spectra from our images using a Python pipeline modeled after \citet{Boogert2002}.  After flat-fielding and dark-subtracting each image, we difference each set of AB pairs.  Going order by order we stack and align the differenced AB pairs and combine them into a single image.  For each order we then fit the spectral trace with a third order polynomial to align the moderately curved 2D spectrum along the \textit{x} (dispersion) axis. 
If the PMO object trace is too faint to get a good fit (as is the case for ROXs12 b, SR12 c, 2M0249-0557 c, OPH 103, and 2M2208+2921), we use a fit to a significantly higher signal-to-noise stellar trace in order to rectify the PMO 2D spectra.  In the case of the bound companions we use fits to their host star spectra, and for the two faint free-floating objects we took trace fits from corresponding standard star observations.  While the star and the PMO were not simultaneously in the slit for observations of 2M0249-0557 c, OPH 103, and 2M2208+2921, the shape of the spectral trace did not significantly change.

The pre-upgrade NIRSPEC detector on occasion exhibited a behavior where one or more sets of every eight rows on the left side of the detector had values that were offset by a constant value.  These offset values were likely caused by variations in bias voltages \citep{Bryan2018}.  While this effect is negligible for high signal-to-noise (S/N) spectra, it becomes significant when the object trace is faint.  We found that observations of ROXs12 b and 2M0249-0557 c both exhibited this distinctive striped pattern on the left half of the detector.  We correct for this effect by computing the median value of unaffected rows and then adding or subtracting a constant value from the offset rows to match this median pixel value.  However, while this process improved the S/N of the spectra on the left side of the detector, we find in subsequent analyses that for both ROXs12 b and 2M0249-0557 c, including the part of the spectrum from the left (blue) half of the detector ultimately degrades the significance of our $v\sin i$ measurements for these objects. We therefore only use the right (red) half of the spectra for these two targets when measuring $v\sin i$.  

\begin{figure}[h]
\centering
\vspace{0.07in}
\includegraphics[width=0.5\textwidth]{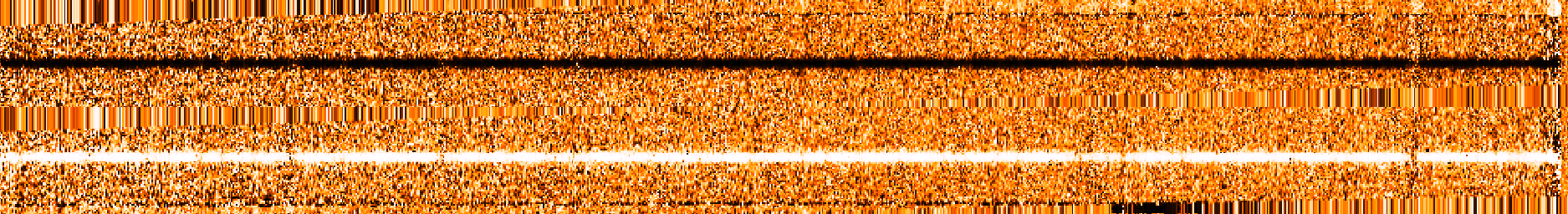} 
\vspace{-0.07in}
\includegraphics[width=0.5\textwidth]{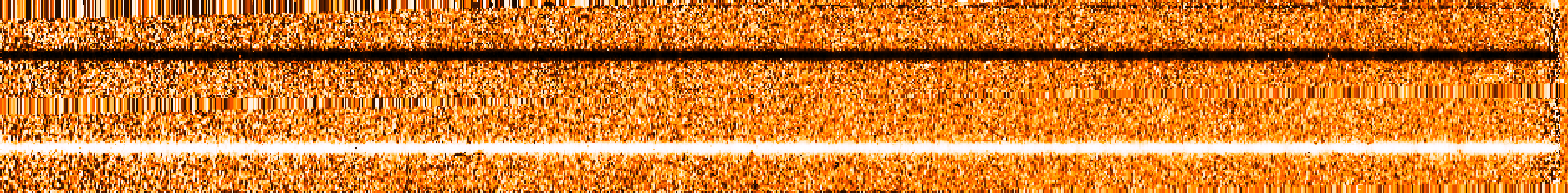} 
\vspace{0.02in}
\caption{Top image: Order 1 2D rectified spectrum for 2M2244+2043 (wavelengths 2.34 - 2.38 $\mu$m).  Bottom image:  Order 2 2D rectified spectrum for 2M2244+2043 (wavelengths 2.27 - 2.31 $\mu$m).  The $y$-axis is the spatial (cross-dispersion) axis, and the $x$-axis is the wavelength (dispersion) axis.  The negative and positive traces in each order come from prior differencing of successive AB pairs before combining.}
\label{fig: 2d spec}
\end{figure}

After producing combined, rectified 2D spectra for each of the six orders (i.e. Fig. \ref{fig: 2d spec}), we optimally extract 1D spectra in pixel space.  For each positive and negative trace, which result from our earlier differencing of AB pairs, we calculate an empirical point spread function (PSF) profile along the $y$ (cross-dispersion) axis of each 2D rectified order.  For the ROXs 12 b and SR 12 c 2D spectra, which include both the stellar trace and the companion trace at the same time, we first plotted the PSF profile of each order to confirm that the light from the star was well separated from the light from the companion.  After identifying the range of $y$ locations of the stellar PSF, we set these to zero.  Finally, we use these PSF profiles to combine the flux of the PMOs along each column, producing 1D spectra in pixel space. 

\subsection{Wavelength Calibration}
To convert these 1D spectra in pixel space to wavelength space, we calculate a wavelength solution for each order.  Since the same instrument configuration (filter, rotator angle, etc.) was maintained over the course of the night, the wavelength solution should remain constant between targets aside from a linear offset due to different placements of targets within the slit.  As a result, we leverage the higher S/N of stellar spectra to calculate a more precise wavelength solution for the fainter PMOs.  For the planetary-mass companions we used their host star spectra, and for the free-floating planetary-mass brown dwarfs we used standard star spectra taken immediately before or after the science targets.  We fit the positions of telluric lines in each stellar spectrum with a third order polynomial wavelength solution of the form $\lambda$ = $ax^3 + bx^2 + cx + d$, where $\lambda$ is wavelength and $x$ is pixel number.  To determine the wavelength solution for the respective PMOs, for each object we apply the stellar wavelength solution to the substellar spectrum, and cross-correlate that spectrum with a telluric model to measure the linear offset.  This offset plus the stellar wavelength solution yields 1D PMO spectra in wavelength space.

\subsection{Telluric Correction}
\begin{figure}[h]
\centering
\includegraphics[width=0.5\textwidth]{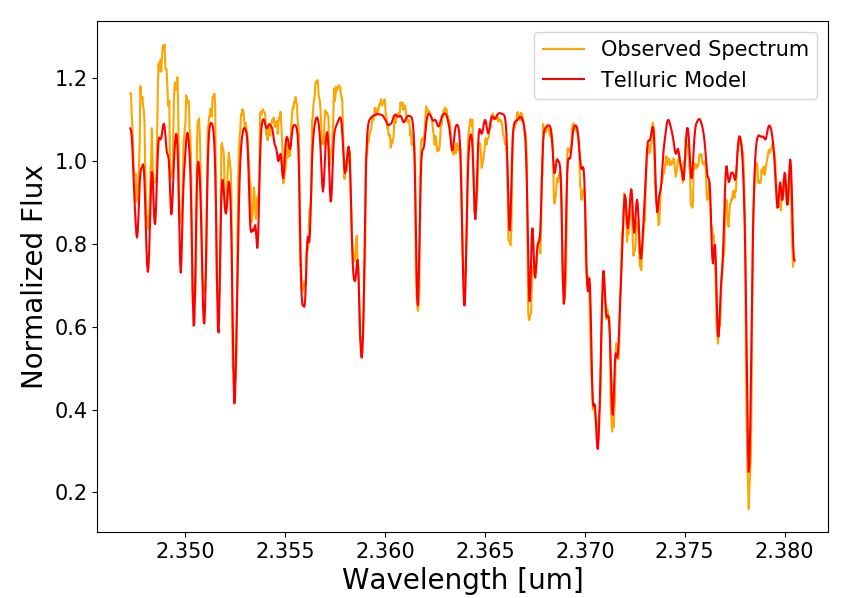} 
\caption{Order 1 negative trace spectrum of standard star HIP 96260 (orange), and the best fit telluric model from \texttt{molecfit} (red).  This telluric model was used to correct the spectrum of 2M2013-2806.}
\label{fig: tell-corr spec}
\end{figure}

We first correct for telluric features in each stellar spectrum using the \texttt{molecfit} routine, which fits a telluric model and an instrumental profile defined by a single Gaussian kernel simultaneously to the spectrum \citep{Smette2015,Kausch2015}.  While we explored an alternative instrumental profile with a central Gaussian and four satellite Gaussians on either side \citep{Valenti1995} in B18, we found that rotational broadening measurements were negligibly impacted between this more complicated kernel and a single Gaussian kernel (conservatively, $v\sin i$ values were consistent at a $<$1$\sigma$ level.  The \texttt{molecfit} routine also uses a third order polynomial to iteratively fit the continuum before dividing out the telluric model.  We use the best fit telluric models for the host and standard stars to telluric-correct the PMO spectra, dividing the relevant telluric model from each PMO spectrum.  Since these telluric corrections are not exact, they leave significant artifacts in the observed spectra where there are strong telluric features due to the fact that deep line cores are difficult to fit well.  As a result we remove these artifacts at the location of the strongest telluric lines.  Figure \ref{fig: tell-corr spec} shows an example 1D wavelength-calibrated spectrum with the best-fit telluric model for order 1 of the standard star HIP 96260 used for telluric-correcting the spectrum of 2M2013-2806.

While there are six orders in pre-upgrade NIRSPEC data, in subsequent analyses we only use orders 1 and 2 of our spectra (wavelength ranges 2.34 - 2.38 $\mu$m and 2.27 - 2.31 $\mu$m respectively).  These two orders at the red end of our spectra have the most accurate wavelength calibrations and telluric corrections, and given the spectral types of the PMOs in our sample these wavelength ranges contain strong and numerous absorption features from both water and CO.

\subsection{Measuring $v \sin i$}
In each spectrum we aim to measure the extent of rotational line broadening due to the spin of the object, which yields $v\sin i$.  For each object, we take the observed wavelength-calibrated, telluric-corrected spectrum and cross correlate it with a model atmosphere that has been broadened to the instrumental resolution.  For all objects except ROXs 12b we use atmospheric models generated using the Sonora model grid \citep[Marley et al. in prep., Morley et al. in prep.]{Marley2018}.  Following the approach of \citet{Marley1999}, \citet{Saumon2008}, and \citet{Morley2012}, the models are calculated assuming the atmosphere is in radiative-convective and chemical equilibrium and using updated chemistry and opacities \citep[Marley et al., in prep.]{Marley2018}.  These models have solar metallicity and C/O ratio, and include iron, silicate, and corundum clouds with a sedimentation efficiency $f_{\rm sed}$ = 2 \citep{Ackerman2001}.  Since the effective temperature of ROXs 12 b (3100 K) is above the T$_{\rm eff}$ limit of the Sonora model grid, we use a BT Settl model for this object.  Table \ref{tb:model param} shows the T$_{\rm eff}$ and log(g) values assumed for each object and used to generate individual models. Assumptions made when generating atmospheric models, such as atmospheric composition, assumed T$_{\rm eff}$ and log(g), etc. can impact the measured rotational velocity by artificially broadening the produced cross-correlation function.  We address these potential biases later in this section.

\begin{deluxetable}{ccc}
\tablecaption{Atmospheric Model Parameters \label{tb:model param}}
\tabletypesize{\footnotesize}
\tablehead{
  \colhead{Object } & 
  \colhead{T$_{\rm eff}$ [K]} & 
  \colhead{log(g) } 
}
\startdata
 ROXs 12 b & 3100 & 4.0 \\
 SR 12 c & 2200 & 3.75 \\
 2M0249-0557 c & 1700 & 4.0 \\
 OPH 98 & 2100 & 3.75  \\
 OPH 103 & 2100 & 3.75  \\
 2M2244+2043 & 1300 & 4.5 \\
 2M2013-2806 & 2100 & 4.0 \\
 2M2208+2921 & 2100 & 4.0
\enddata
\end{deluxetable}

We next compare the shape of the observed cross correlation functions (CCF) for each object to ``model'' CCFs.  Each model CCF is calculated by taking a model atmosphere generated with the parameters from Table \ref{tb:model param}, and cross-correlating it with the same model additionally broadened by some rotation rate and shifted by a radial velocity.  We quantify this comparison in a Bayesian framework using MCMC, fitting for $v\sin i$, a radial velocity offset, and instrumental resolution.  While instrumental resolution is degenerate with $v\sin i$ -- both serve to broaden the absorption line profile -- we have empirical measurements of NIRSPEC instrumental resolution from calculations of optimal telluric models using \texttt{molecfit}.   We can thus compute an error weighted average instrumental resolution for each night that we observed; when we examine individual spectra, we find that this resolution remains stable over the course of a night.  In the MCMC framework, we adopt a Gaussian prior for instrumental resolution, with a peak location and width equal to the error-weighted average instrumental resolution and uncertainty associated with the observation date.  We assume uniform priors for both $v\sin i$ and radial velocity (RV).  In \citet{Bryan2020} we tested whether our assumption of a uniform prior on $v\sin i$ could bias our rotation rate measurements.  We split up $v\sin i$ into $v$ and $i$, assuming a uniform prior for $v$ and a uniform prior in $\cos i$ for $i$, and found that the measured $v\sin i$ differed from the original by $<0.2\sigma$.  

Equation \ref{eq:loglike} shows the log likelihood function we employed in our framework:

\begin{equation}
\log{L} = \sum_{i=1}^{n} -0.5\bigg(\frac{m_i - d_i}{\sigma_{i}}\bigg)^2,
\label{eq:loglike}
\end{equation}
\noindent where $d$ is the observed CCF calculated by cross correlating an observed spectrum with the corresponding model atmosphere broadened to the instrumental resolution, and $m$ is the model CCF calculated by cross correlating a model atmosphere broadened to the instrumental resolution with that same model additionally broadened by a $v\sin i$ value and offset by an RV.  We drop the error term in the log likelihood equation (-0.5$\log(2\pi \sigma^2)$) because it is constant in our fits.  We calculate uncertainties on the observed CCF $\sigma$ using the jackknife resampling technique defined by equation \ref{eq:jackknife}:

\begin{equation}
\sigma_{\rm{jackknife}}^2 = \frac{(n-1)}{n} \sum_{i=1}^{n} {(x_i - x)}^2,
\label{eq:jackknife}
\end{equation}
\noindent where $n$ here is the total number of AB pairs in an object's dataset, $x_i$ is the CCF calculated with all but the $i^{th}$ AB pair, and $x$ is the observed CCF calculated using all AB pairs.  The number of nod pairs varied between three and twelve in our sample (see Table \ref{tb:obs} for details).  

Apart from the calculated uncertainties on the observed CCFs, we check whether measured $v\sin i$ values could be artificially inflated by small offsets in wavelength space between individual AB nod spectra.  We test this by calculating CCFs for each individual AB pair, treating the positive and negative traces separately, and measuring the locations of the CCF peaks.  For traces that exhibit significant CCF peaks, we find that for 2M2013-2806, 2M2208+2921, and 2M2244+2043, the relative spread in the individual CCF peaks was more than 5 km/s.  We thus shift the wavelengths of each individual AB pair spectrum according to the measured CCF peak offsets.  After these shifts, for our complete sample we found that within the positive trace and negative trace spectra, the observed wavelength shifts were small, typically less than 2 km/s, and thus would not artificially inflate measured $v\sin i$.  For each positive and negative trace we combine individual AB pair spectra before using them to measure $v\sin i$.  

\begin{figure}[h]
\centering
\includegraphics[width=0.5\textwidth]{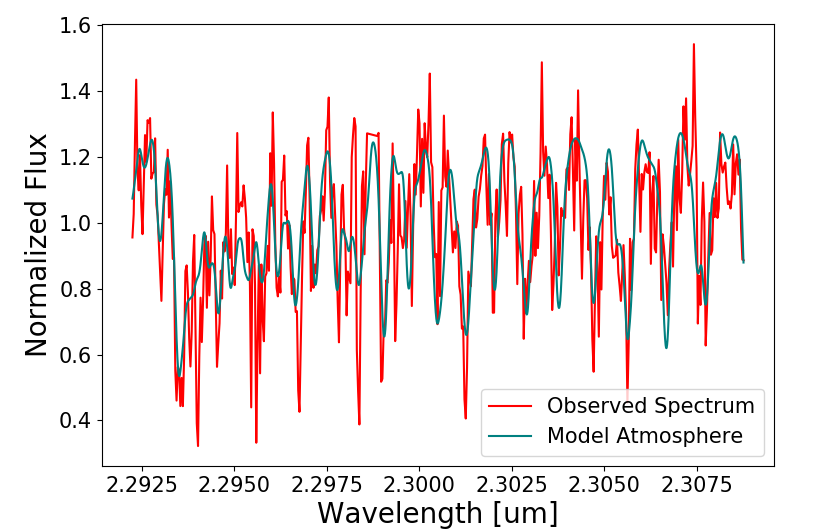} 
\caption{Observed spectrum for 2M0249-0557 c (red), and the best-fit atmospheric model (teal).  The model has been broadened by the measured instrumental resolution as well as the best-fit $v\sin i$ value, and shifted to the best-fit radial velocity offset. }
\label{fig: PMO spectra}
\end{figure}

\begin{figure*}
    \epsscale{1.1}
    \centering
    \plottwo{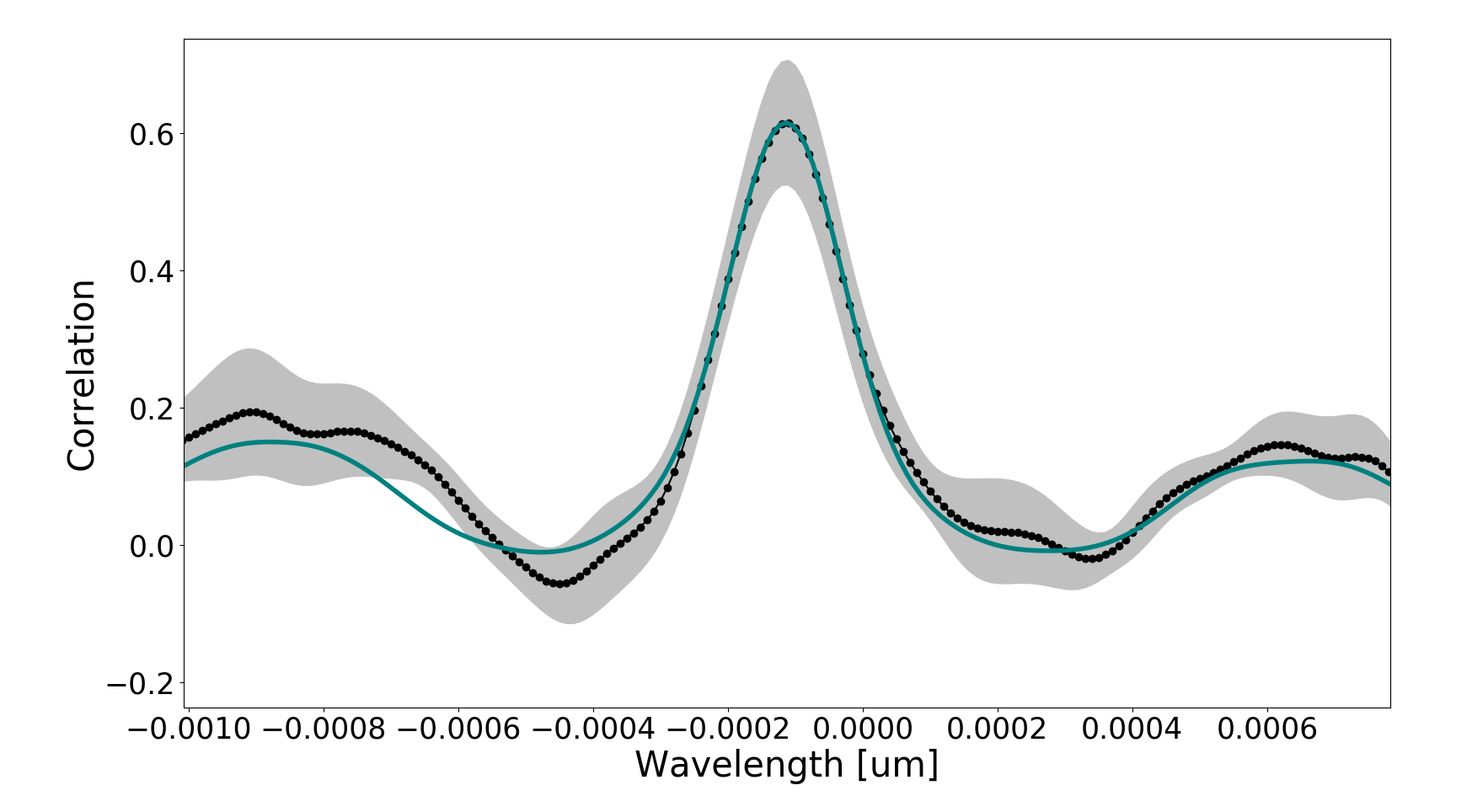}{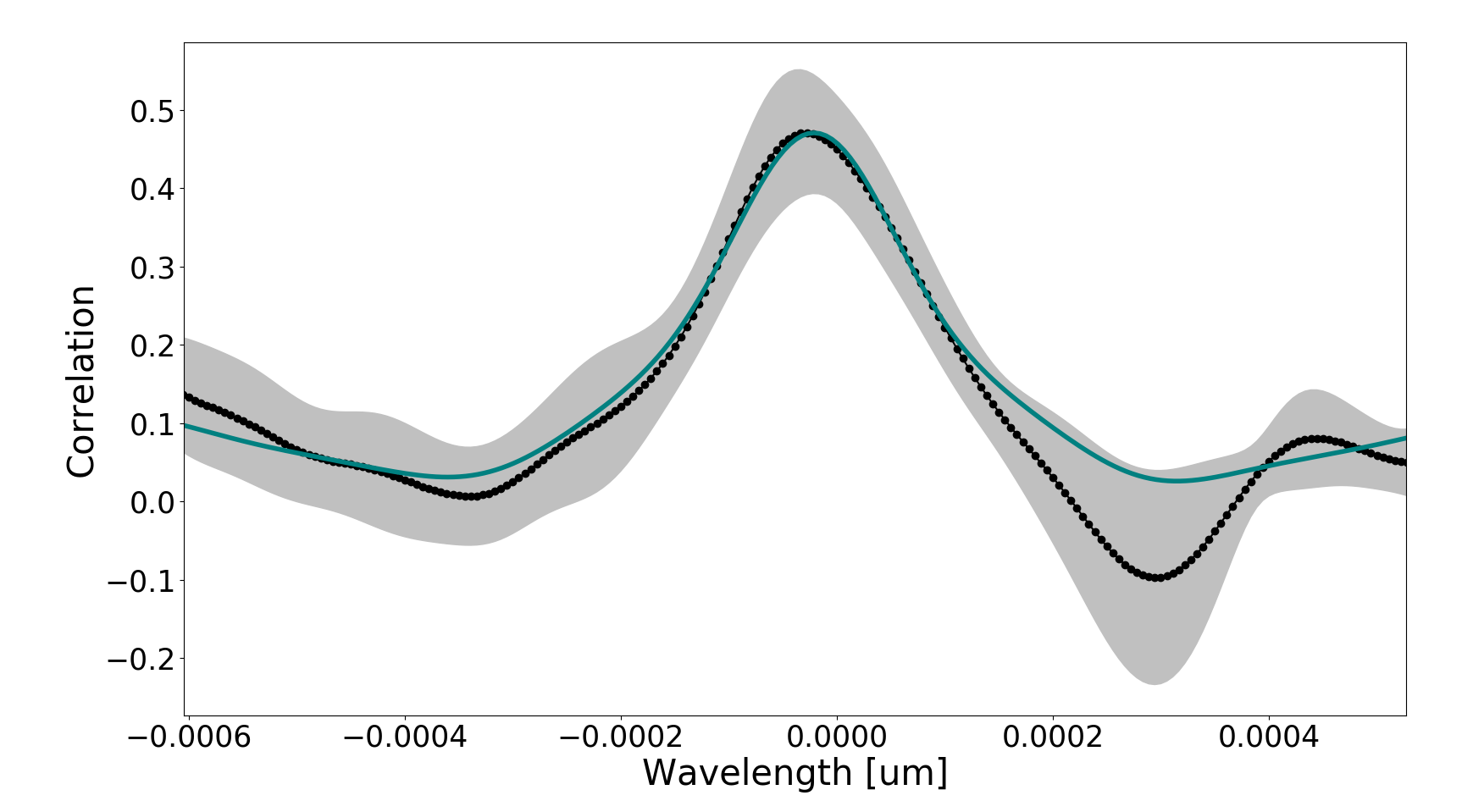} \\
    \plottwo{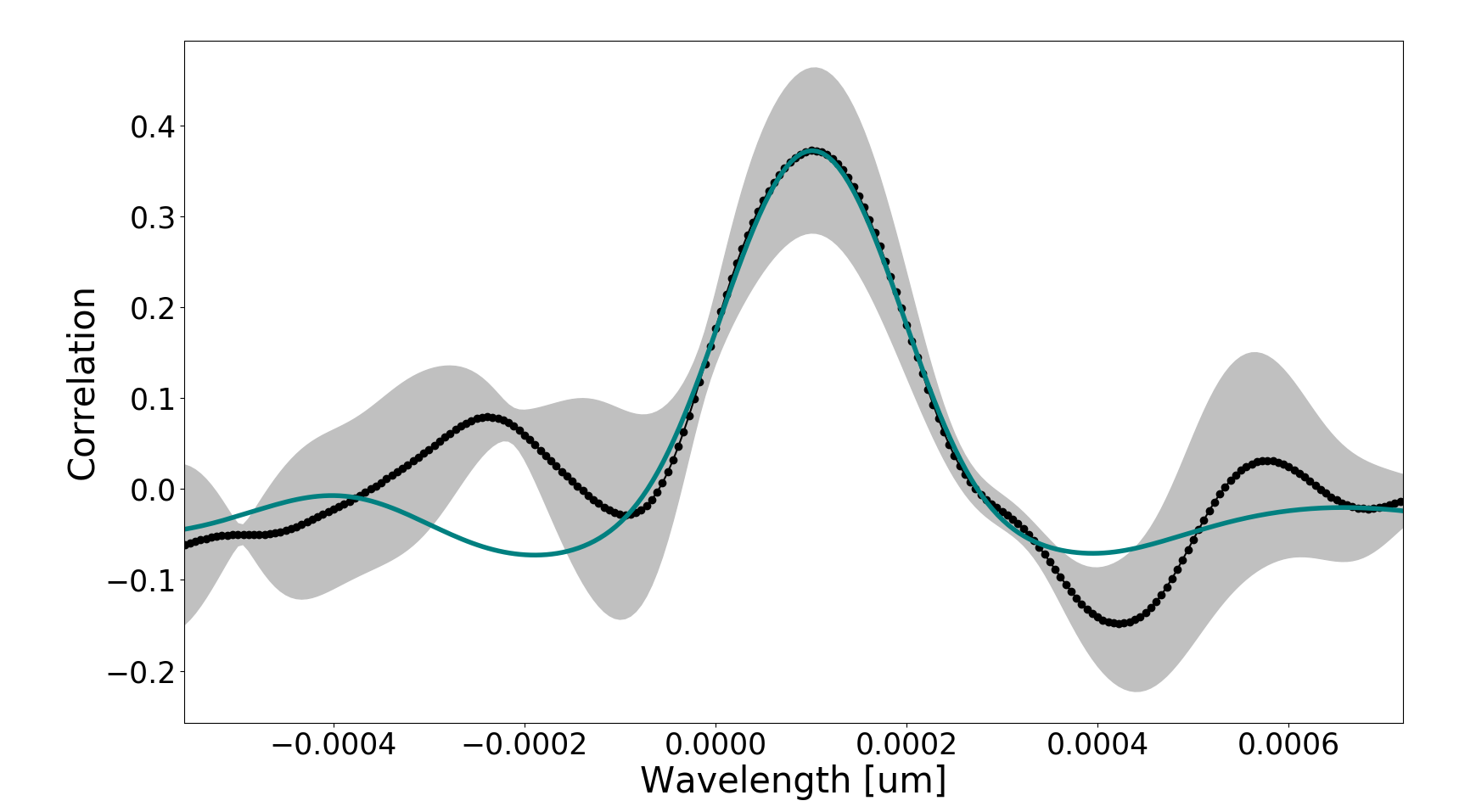}{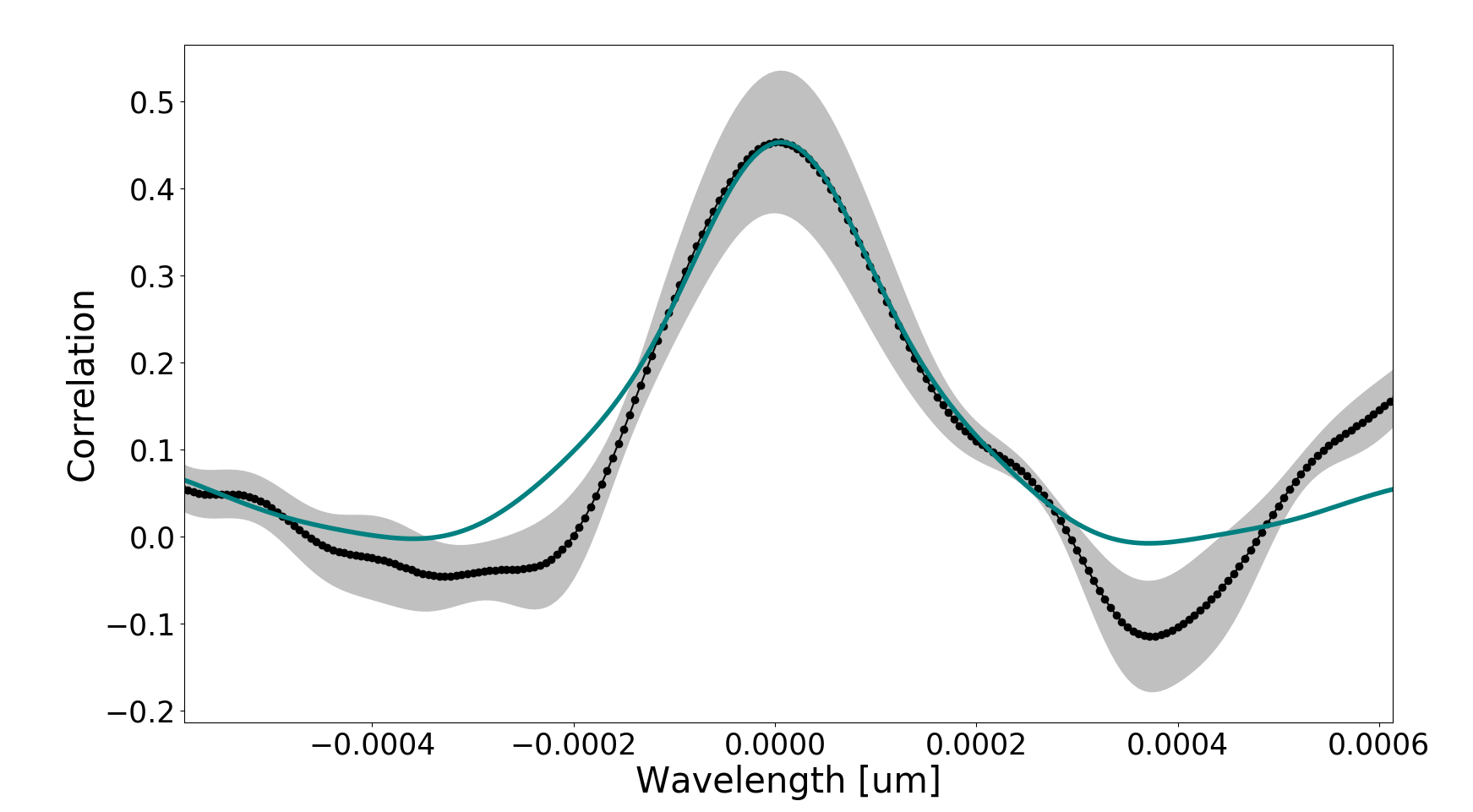} \\
    \plottwo{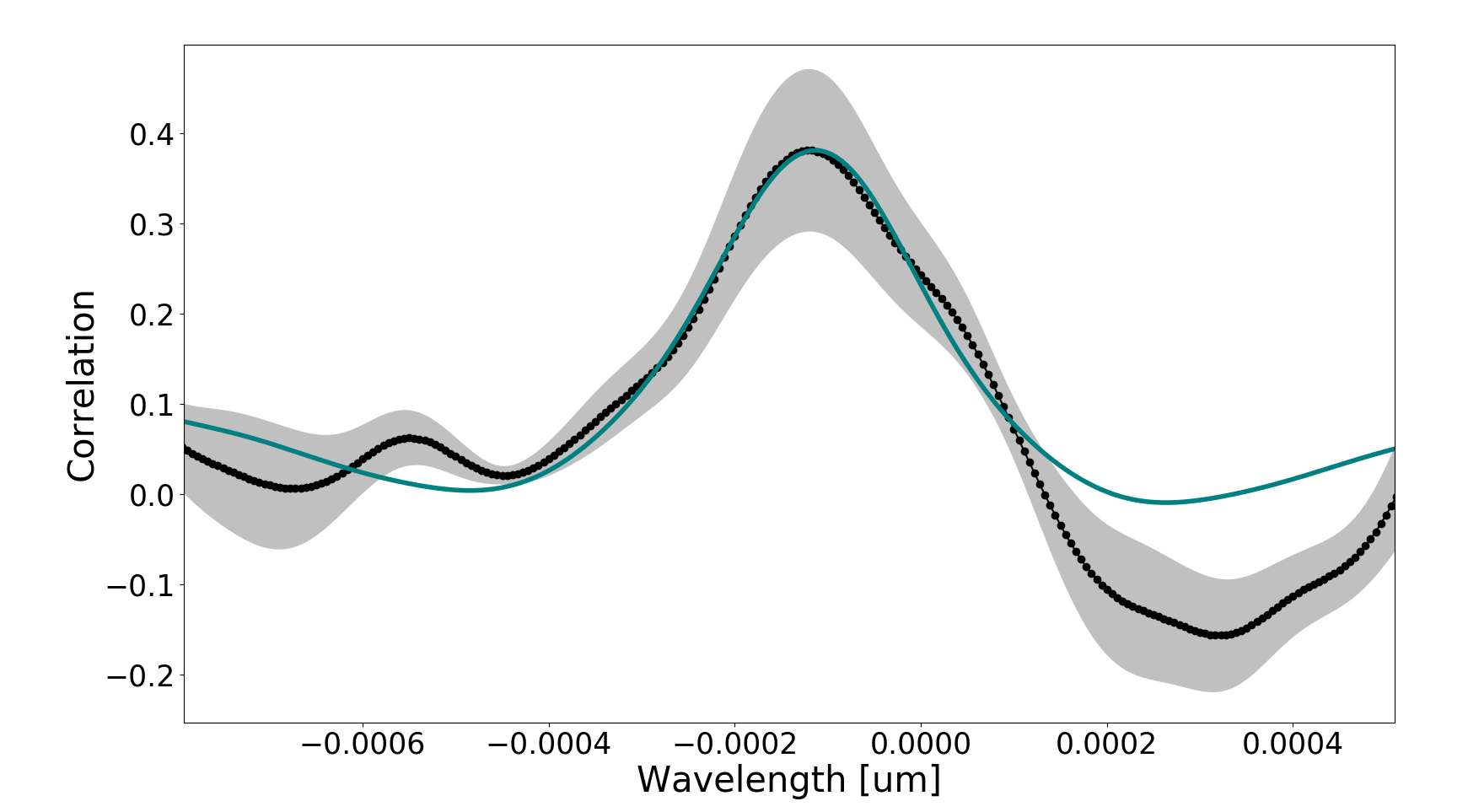}{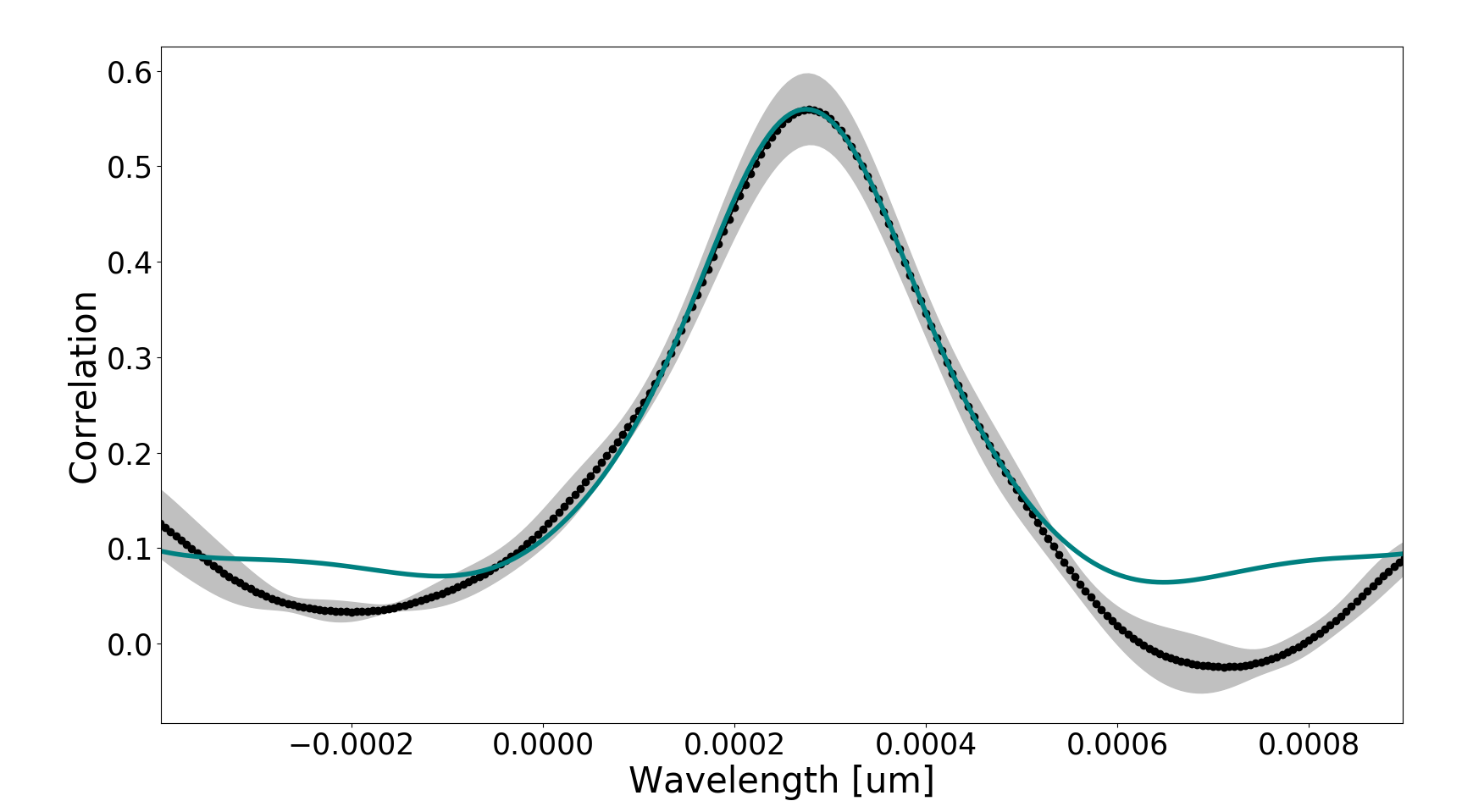} \\
    \plottwo{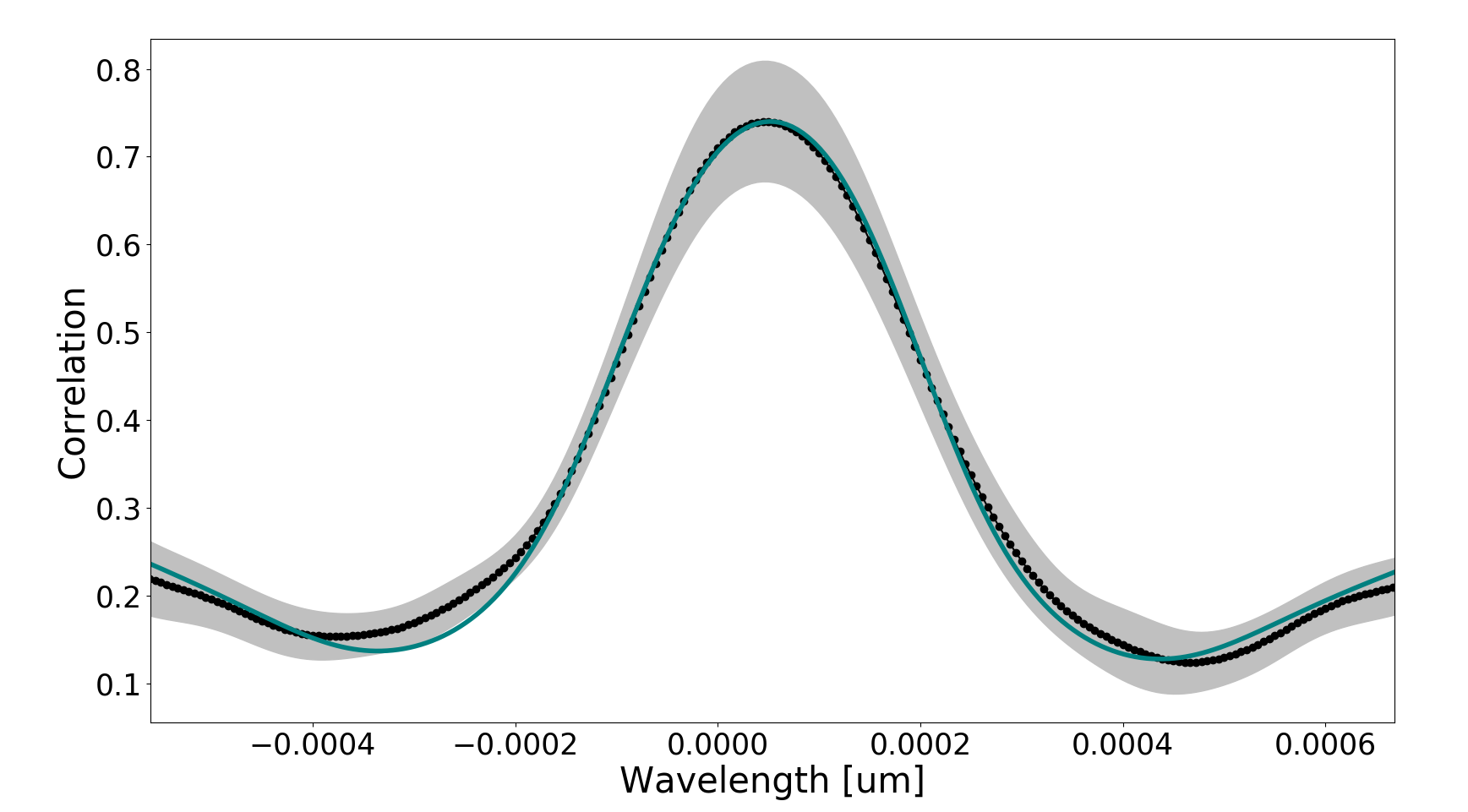}{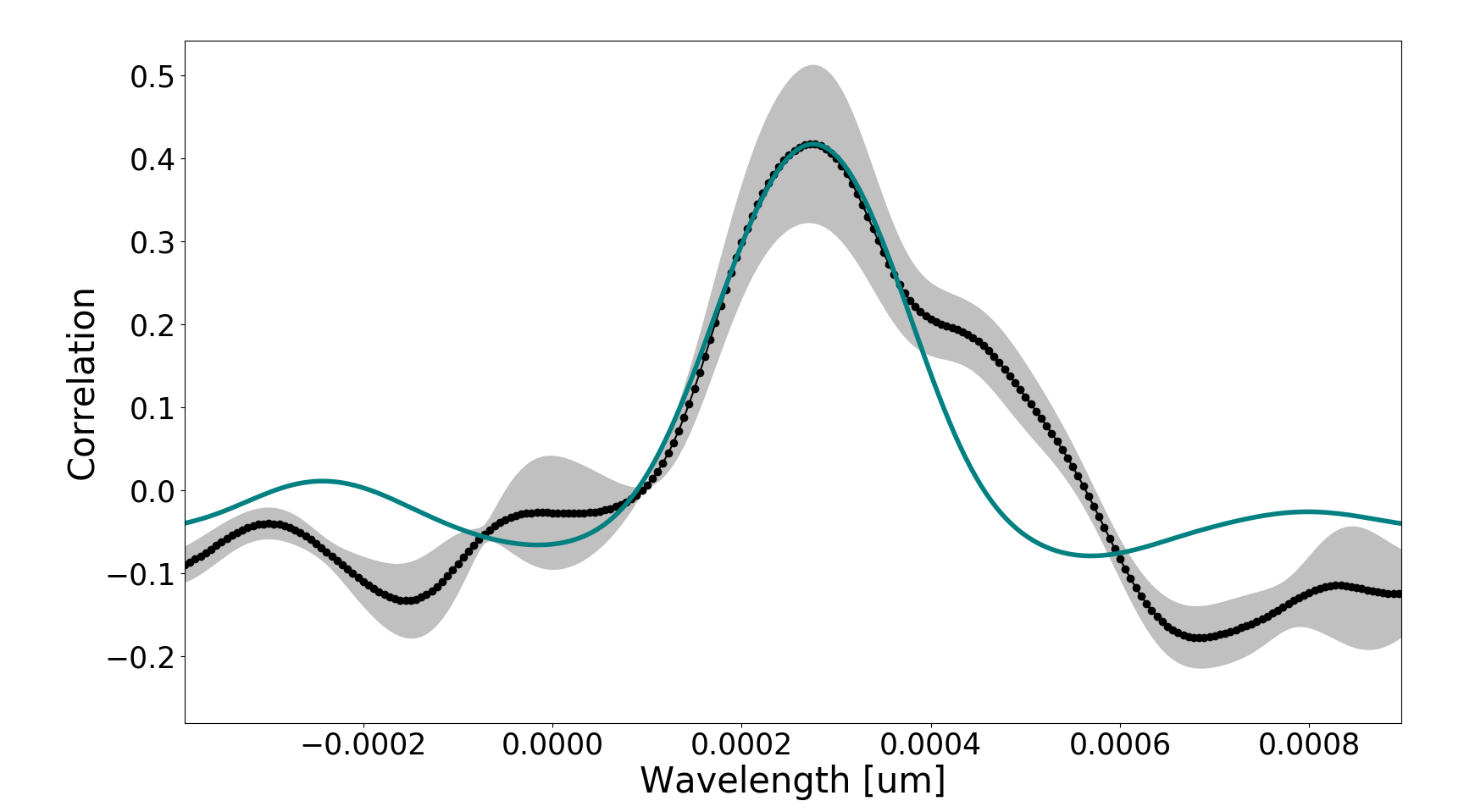} 
    \caption{Cross correlation functions between observed spectra for each PMO with respective model atmospheres broadened to the instrumental resolution (black), with 1$\sigma$ uncertainties from the jackknife resampling technique shaded in gray.  Overplotted in teal for each object is the model CCF calculated using an atmospheric model broadened to the instrumental resolution, with that same model additionally broadened by the best-fit $v\sin i$ value and velocity offset.  In order, the objects are:  ROXs 12 b, SR 12 c, 2M0249-0557 c, OPH 98, OPH 103, 2M2244+2043, 2M2013-2806, and 2M2208+2921.
    \label{fig:PMO CCFs}}
\end{figure*}

We leave the positive and negative trace spectra separate for our sample both because the wavelength offsets between the two are significant for all systems and also because fitting both traces for each order provides independent estimates of $v\sin i$.  In addition, we fit $v\sin i$ separately for orders 1 and 2.  However, some traces had a S/N that was too low to produce a significant CCF peak, meaning we could not measure a $v\sin i$ value.  These spectra are:  SR 12 c order 1 negative trace, OPH 98 order 1 positive and negative trace, OPH 103 order 1 positive and negative trace, and 2M2208+2921 order 1 negative trace.  We found that the significance could vary between positive and negative traces based on how well the PMO is centered in the slit (each trace corresponds to an A or B nod).  For all objects in our sample, individual measured $v\sin i$ values are consistent within their uncertainties to $<$2$\sigma$.  We calculate their error weighted averages, shown in Table \ref{tb:res} (Figs. \ref{fig: PMO spectra}, \ref{fig:PMO CCFs}).

Measured $v\sin i$ values can also be impacted by assumptions made when generating atmospheric models.  In \citet{Bryan2020}, we used the spectrum of the planetary-mass companion 2M0122-2439 b and tested our choice of T$_{\rm eff}$ and log(g), C/O ratio, and pressure broadening on the resulting $v\sin i$ measurement.  We first produced atmospheric models calculated using T$_{\rm eff}$ and log(g) values 1$\sigma$ away from the best fit values, and found that the resulting rotational velocity differed from the original by $<0.7\sigma$.  Next, we produced models with 0.25$\times$solar, 0.5$\times$solar, and 1.5$\times$solar C/O ratios and concluded that resulting $v\sin i$ values were consistent with the original at the 0.8$\sigma$ level.  Finally, in \citet{Bryan2020} we investigated the effect of pressure broadening uncertainties on the measured 2M0122-2439 b rotation rate by simulating scenarios where pressure broadening parameters used to create molecular cross sections are off by an order of magnitude in each direction, and determined that the new values differed from the original by less than 0.6$\sigma$.  For all of these cases, assumptions made when producing models do not significantly impact the resulting $v\sin i$ measurements.  The PMOs presented here have comparable parameters (such as mass, spectral type, temperature) to 2M0122-2439 b as well as to PMOs in B18 where similar tests were run. We conclude that our new sample of PMOs will be similarly unaffected by assumptions regarding composition, pressure broadening, and $T_{\rm eff}$ and log(g).  

The new spin measurements span a range of $v\sin i$ from 8.4 -- 21.8 km/s, consistent with published rotational velocities for other objects in the planetary-mass regime (Table \ref{tb:res}).  These new measurements are the first rotation rate measurements made for this sample of eight PMOs.  We combine our sample of eight new $v\sin i$ measurements with 19 published rotational velocities.  Eight of these published values are $v\sin i$ measurements from our NIRSPEC/Keck survey \citep{Bryan2018,Xuan2020,Bryan2020}, and five more are $v\sin i$ measurements from other surveys \citep{Snellen2014,Crossfield2014,Mohanty2005,Gagne2015}.  The remaining published rotation rates are photometric rotation periods from measured light curve variability \citep{Zhou2016,Zhou2019,Zhou2020,Manjavacas2019,Miles-Paez2019,Schneider2018}.  Two of these objects, namely 2M0122-2439 b and VHS 1256-1257 b, have both measured $v\sin i$ and measured $P_{\rm rot}$.  Combined, these measurements yield the line-of-sight inclination of the PMO's spin axis.  \citet{Bryan2020} combined this inclination with measurements of the orbital inclination to constrain the obliquity of 2M0122-2439 b.

\subsection{Calculating PMO Parameters}

In subsequent analyses we consider how rotation rates in the planetary-mass regime correlate with properties such as object radius, semi-major axis, and system mass ratio.  While the bound PMOs have previous mass estimates (see Table \ref{tb:res} and references therein), we derive masses for the free-floating PMOs.  First we take object distance, spectral type, and K-band bolometric correction for young ultracool dwarfs from \citet{Filippazzo2015} to compute each bolometric luminosity.  This luminosity along with object age allows us to compute object mass $M$ using a finely interpolated hot-start evolutionary model grid \citep{Burrows1997}.  Uncertainties on $M$ are computed by incorporating uncertainties on apparent K-band magnitude, distance, and spectral type in a Monte Carlo fashion.  For 5$\times$10$^5$ trials we draw each of these three values randomly from a Gaussian distribution whose peak and width are defined by the best-fit value and uncertainty of the respective variable, and propagate these values to create a probability distribution for $M$.  We note that while we assume hot-start models here, post-formation luminosities could be between cold and hot start models \citep{Mordasini2017}, and recent work indicates that accretion likely results in either warm or hot starts \citep{Berardo2017,Berardo22017,Marleau2019}.  Assuming lower post-formation entropies (as is the case for cold or warm start relative to hot start) yields higher mass estimates \citep{Marleau2014}.  Because all mass estimates considered in this paper were calculated using hot-start models, true masses could be higher than values given in Table \ref{tb:res}.  However, this uncertainty does not impact our findings in section 4 since objects in our sample have comparable masses.

We estimate radii for all PMOs in our sample in a similar fashion.  Given an age and bolometric luminosity for each object, we interpolate an evolutionary model grid and use the Monte Carlo sampling approach to produce probability distributions for each radius.  The masses and radii used in subsequent analyses are shown in Table \ref{tb:res}.  

Since all bound objects are directly imaged, they have directly measured projected separations, initially in arcseconds but simply converted to AU using known distances.  For our purposes we want to convert these projected separations to semi-major axes.  Again employing the Monte Carlo sampling method, we randomly generate values for companion orbital phase, inclination, eccentricity, argument of periastron, and longitude of ascending node.  We draw orbital phase values from a uniform distribution from 0 to 1, argument of peristron and longitude of ascending node from uniform distributions from 0 to 2$\pi$, inclination from a uniform distribution in $\sin i$, and adopt an eccentricity of 0 \citep[i.e.][]{Bowler2015}.  Combined with the measured projected separation distributions, we compute semi-major axis $a$ values for each set of drawn orbital parameters, producing a distribution for $a$.  Table \ref{tb:res} shows resulting best-fit $a$ values and uncertainties.

Using stellar mass estimates from the literature (Table \ref{tb:res}), we calculate system mass ratios $M/M_{\rm star}$.  We analytically propagate uncertainties as:

\begin{equation}
\sigma_{\rm ratio} = \frac{M}{M_{\rm star}}\sqrt{\bigg(\frac{\sigma_M}{M}\bigg)^2 + \bigg(\frac{\sigma_{\rm Mstar}}{M_{\rm star}}\bigg)^2}
\label{eq:sigma_ratio}
\end{equation}

Before investigating trends between PMO spin and other system parameters, we convert all direct measurements of spin, both $v\sin i$ and $P_{\rm rot}$, to equatorial rotation rate $v$ [km/s].  For each $v\sin i$ measurement we divide that probability distribution by a distribution of $\sin i$, where inclinations are drawn from a uniform distribution in $\cos i$.  We convert all $P_{\rm rot}$ measurements to $v$ using $v$ = 2$\pi$$R$/$P_{\rm rot}$.  We note that while brown dwarf companions GQ Lup B and HN Peg B and free-floating brown dwarfs such as 2MASS J0501-00, 2MASS J0045+16, and 2MASS J1425-36 have measured rotation speeds

\clearpage
\begin{turnpage}
\begin{deluxetable}{lccccccccccc}
\tablecaption{Physical and rotational properties of planetary-mass objects \label{tb:res}}
\tabletypesize{\footnotesize}
\tablehead{
  \colhead{Name } & 
  \colhead{v$\sin$i } & 
  \colhead{P$_{\rm rot}$ } & 
  \colhead{v} & 
  \colhead{v/v$_{\rm break}$ } & 
  \colhead{Spin source} & 
  \colhead{M$_{\rm pl}$} &
  \colhead{R$_{\rm pl}$} &
  \colhead{Age} &
  \colhead{a} &
  \colhead{M$_{\star}$\tablenotemark{a}} &
  \colhead{References}\\
  \colhead{} & 
  \colhead{[km/s]} & 
  \colhead{[hrs]} & 
  \colhead{[km/s]} & 
  \colhead{} & 
  \colhead{} &
  \colhead{[M$_{\rm Jup}$]} &
  \colhead{[R$_{\rm Jup}$]} &
  \colhead{[Myr]} &
  \colhead{[AU]} &
  \colhead{[M$_{\odot}$]} &
  \colhead{}
}
\startdata
\cutinhead{Bound Planetary-Mass Companions}
 \vspace{-0.1in}\\

 ROXs 12 b& 8.4$^{+2.1}_{-1.4}$ & $\cdots$ & 10.4$^{+2.7}_{-4.4}$ & 0.09$^{+0.02}_{-0.04}$ & This paper & 17.5$\pm$1.5 & 2.31$^{+0.10}_{-0.15}$ & 6$^{+4}_{-2}$ & 236$^{+41}_{-50}$ & 0.65$^{+0.05}_{-0.09}$  & 1, 15, 16\\
 \vspace{-0.1in}\\
 SR 12 c & 9.7$^{+1.2}_{-1.4}$ & $\cdots$ & 11.5$^{+2.4}_{-3.5}$ & 0.12$\pm$0.03 & This paper & 13$\pm$2 & 2.38$^{+0.32}_{-0.27}$ & 3$\pm$2 & 1208$^{+273}_{-354}$ & 1.05$\pm$0.05, 0.5$\pm$0.1  & 1, 17, 18, 19\\
 \vspace{-0.1in}\\
 2M0249-0557 c &  15.3$^{+0.4}_{-0.7}$ & $\cdots$ & 17.8$^{+3.6}_{-3.0}$ & 0.15$\pm$0.03 & This paper & 11.6$^{+1.3}_{-1.0}$ & 1.43$\pm$0.03 & 22$\pm$6 & 2156$^{+354}_{-488}$ &  0.046$\pm$0.012,0.042$^{+0.013}_{-0.010}$  & 1, 20\\
 \vspace{-0.1in}\\
 2M0122-2439 b & 13.4$^{+1.4}_{-1.2}$ & 6.0$^{+2.6}_{-1.0}$ & 15.9$^{+3.0}_{-4.5}$ & 0.09$^{+0.02}_{-0.03}$ & Bryan+20/Zhou+19 & 12$-$27 & 1.17$\pm$0.02 & 120$\pm$10 & 59$^{+9}_{-14}$  &  0.40$\pm$0.05  & 1, 2, 21\\
 \vspace{-0.1in}\\
 DH Tau b & 9.6$\pm$0.7 &$\cdots$ & 11.2$^{+2.4}_{-2.8}$ & 0.11$^{+0.03}_{-0.04}$  & Xuan+20 & 8$-$22 & 2.68$\pm$0.22 & 2$\pm$1 & 345$^{+53}_{-41}$ & 0.64$\pm$0.05 &  1, 3, 16, 22, 23\\
 \vspace{-0.1in}\\
 ROXs 42B b & 9.5$^{+2.1}_{-2.3}$ & $\cdots$ & 11.7$^{+3.1}_{-4.2}$ & 0.12$^{+0.04}_{-0.05}$ & Bryan+18 & 10$\pm$4 & 2.11$\pm$0.11 & 3$\pm$2 & 153$^{+21}_{-27}$ & 0.89$\pm$0.08 &  1, 4, 16, 17, 24\\
 \vspace{-0.1in}\\
 VHS 1256-1257 b & 13.5$^{+3.6}_{-4.1}$ & 22.04$\pm$0.05 & 16.6$^{+5.8}_{-7.0}$ & 0.12$^{+0.07}_{-0.05}$ & Bryan+18/Zhou+20 & 11.2$^{+9.7}_{-1.8}$ & 1.11$\pm$0.03 & 150$-$300 & 112$^{+12}_{-26}$ & 0.0616$^{+0.0008}_{-0.0019}$,0.616$^{+0.0008}_{-0.0019}$   &  1, 4, 25, 26\\
 \vspace{-0.1in}\\
 GSC 6214-210 b & 6.1$^{+4.9}_{-3.8}$ & $\cdots$ & 7.7$^{5.3}_{-5.7}$ & 0.07$^{+0.03}_{-0.06}$ & Bryan+18 & 15$\pm$2 & 1.91$\pm$0.07 &  11$\pm$2 & 354$^{+57}_{-76}$  & 0.9$\pm$0.1 &  1, 4, 27, 28\\
 \vspace{-0.1in}\\
 $\beta$ Pic b & 25.0$\pm$3.0 & $\cdots$ & 29.7$^{+6.1}_{-8.8}$ & 0.24$^{+0.05}_{-0.07}$ & Snellen+14 & 13$\pm$3 & 1.47$\pm$0.02 & 22$\pm$6 & 11.8$^{+0.8}_{-0.9}$\tablenotemark{b} &  1.84$\pm$0.05  &  1, 5, 29, 30\\
 \vspace{-0.1in}\\
 2M1207-3932 b & $\cdots$ & 10.7$^{+1.2}_{-0.6}$ & 16.1$^{+1.8}_{-0.9}$ & 0.20$^{+0.05}_{-0.04}$ & Zhou+16 & 5$\pm$2 & 1.38$\pm$0.02 &  10$\pm$3 & 44$^{+6}_{-3}$ & 0.0190$^{+0.0009}_{-0.0012}$ &  1, 4, 6, 31, 32\\
 \vspace{-0.1in}\\
 AB Pic b & $\cdots$ & 2.12$\pm$0.05 & 83.2$\pm$3.6 & 0.66$\pm$0.05 & Zhou+19 & 11$-$14 & 1.40$\pm$0.05 & 15$-$40 & 291$^{+35}_{-19}$ & 0.95 &  1, 7, 33\\
 \vspace{-0.1in}\\
 ROSS 458 c & $\cdots$ & 6.8$\pm$1.6 & 19.8$\pm$4.6 & 0.16$\pm$0.05 & Manjavacas+19 & 9$\pm$3 & 1.07$\pm$0.02 & 150$-$800 & 1275$^{+174}_{-85}$ & 0.6,0.09 & 1, 8, 34, 35\\
 \vspace{-0.1in} \\
 HD 203030 b & $\cdots$ & 7.5$^{+0.6}_{-0.5}$ & 19.8$^{+1.7}_{-1.4}$ & 0.15$\pm$0.03 & Miles-Paez+19 & 8$-$15 & 1.19$\pm$0.03 & 30$-$150 & 516$^{+60}_{-32}$ & 0.95 &  1, 9, 36, 37\\
 \vspace{-0.1in}\\
 HD 106906 b & $\cdots$ & 4.0$\pm$0.3 & 48.7$^{+3.8}_{-3.9}$ & 0.44$\pm$0.05 & Zhou+20 & 11$\pm$2 & 1.56$^{+0.04}_{-0.05}$ & 15$\pm$3 & 786$^{+100}_{-60}$ & 1.5$\pm$0.1 &  1, 10, 38\\
 \vspace{-0.1in}\\
 \cutinhead{Free-floating Planetary-Mass Objects}
 \vspace{-0.1in}\\
 OPH 98 & 14.4$^{+1.2}_{-1.1}$ & $\cdots$ & 16.9$^{+3.3}_{-4.3}$  & 0.17$\pm$0.04 & This paper & 10.9$\pm$2.3 & 1.95$^{+0.11}_{-0.10}$ & 3$\pm$2 & $\cdots$ & $\cdots$ &  1, 39\\
 \vspace{-0.1in}\\
 OPH 103 & 15.3$^{+1.3}_{-3.2}$ & $\cdots$ & 18.2$^{+3.8}_{-4.8}$ & 0.18$^{+0.04}_{-0.05}$ & This paper & 10.4$\pm$2.3 & 1.92$^{+0.11}_{-0.10}$ & 3$\pm$2 & $\cdots$ & $\cdots$ & 1, 39 \\
 \vspace{-0.1in}\\
 2M2244+2043 & 12.4$\pm$0.6 &$\cdots$ & 14.5$^{+3.1}_{-2.8}$ & 0.09$\pm$0.02 & This paper & 16.9$\pm$2.0 & 1.20$^{+0.02}_{-0.01}$ & 120$\pm$10 &$\cdots$ & $\cdots$ &  1, 40 \\
 \vspace{-0.1in}\\
 2M2013-2806 & 21.8$\pm$0.4 &$\cdots$ & 25.3$^{+4.9}_{-3.9}$ & 0.19$^{+0.04}_{-0.03}$  & This paper & 14.2$\pm$1.2 & 1.51$^{+0.04}_{-0.06}$ & 22$\pm$6 & $\cdots$ & $\cdots$ & 1, 41 \\
 \vspace{-0.1in}\\
 2M2208+2921 & 16.1$\pm$1.3 &$\cdots$ & 18.9$^{+3.8}_{-4.6}$ & 0.15$^{+0.03}_{-0.04}$ & This paper & 13.7$\pm$1.2 & 1.51$^{+0.04}_{-0.05}$ & 22$\pm$6 & $\cdots$ & $\cdots$ & 1, 20 \\
 \vspace{-0.1in}\\
 OPH 90 & 8.4$^{+5.5}_{-5.0}$ & $\cdots$ & 10.8$^{+8.2}_{-7.4}$ & 0.11$^{+0.07}_{-0.08}$ & Bryan+18 & 11$\pm$2 & 2.00$^{+0.09}_{-0.12}$ & 3$\pm$2 & $\cdots$ &$\cdots$  & 1, 4, 39 \\
 \vspace{-0.1in}\\
 USco 1608-2315 & 16.3$^{+2.4}_{-2.5}$ & $\cdots$ & 19.8$^{+4.0}_{-6.5}$ & 0.16$^{+0.03}_{-0.05}$ & Bryan+18 & 19$\pm$1.5 & 2.36$^{+0.16}_{-0.21}$ & 11$\pm$2 & $\cdots$ & $\cdots$ & 1, 4, 42 \\
 \vspace{-0.1in}\\
 PSO J318.5-22 & 12.0$^{+3.5}_{-4.4}$ & $\cdots$ & 14.8$^{+4.8}_{-7.2}$ & 0.14$^{+0.04}_{-0.06}$ & Bryan+18 & 8.3$\pm$0.5 & 1.33$\pm$0.02 & 21$\pm$4 & $\cdots$ & $\cdots$ & 1, 4, 43 \\
 \vspace{-0.1in}\\
 2M1207-3932 & 13.7$\pm$1.9 & $\cdots$ & 16.6$^{+3.3}_{-6.0}$ & 0.14$^{+0.03}_{-0.04}$ & Crossfield+14 & 19.9$^{+0.9}_{-1.2}$ & 2.47$\pm$0.12 & 10$\pm$3 & $\cdots$ & $\cdots$ & 1, 11, 44 \\
 \vspace{-0.1in}\\
 GY 141 & 4.9$^\pm$1.1 & $\cdots$ & 6.0$^{+1.8}_{-2.3}$ & 0.05$\pm$0.02 & Crossfield+14 & 17.9$^{+1.9}_{-2.0}$ & 2.58$^{+0.13}_{-0.21}$ & 3$\pm$2 & $\cdots$ & $\cdots$ &  1, 11, 45 \\
 \vspace{-0.1in}\\
 KPNO Tau 12 & 5.0$\pm$2.0 & $\cdots$ & 6.1$^{+2.7}_{-3.4}$ & 0.06$\pm$0.03 & Mohanty+05 & 12.7$^{+1.6}_{-1.8}$ & 2.22$^{+0.11}_{-0.17}$ &  2$\pm$1 & $\cdots$ &$\cdots$  & 1, 12 \\
 \vspace{-0.1in}\\
 SIMP 0136 & 50.9$\pm$0.8 & $\cdots$ & 59.2$^{+11.0}_{-10.0}$ & 0.42$^{+0.08}_{-0.06}$ & Gagne+17 & 12.7$\pm$1.0 & 1.14$\pm$0.04 & 200$\pm$50 & $\cdots$ & $\cdots$ & 1, 13 \\
 \vspace{-0.1in}\\
 WISE 1147-2040 & $\cdots$ & 19.4$\pm$0.3 & 9.1$\pm$0.3 & 0.10$\pm$0.01 & Schneider+18 & 6.6$\pm$1.9 & 1.42$^{+0.03}_{-0.04}$ & 5$-$15 & $\cdots$ & $\cdots$ & 1, 14, 46
\enddata
\tablenotetext{a}{For systems where stellar masses do not have published uncertainties, we assume a conservative error of 0.1 M$_{\odot}$.  }

\tablenotetext{b}{For $\beta$ Pic b, the semi-major axis comes from a fit to measured orbital motion instead of a conversion from projected separation. See section 3.5 for details.}

\tablerefs{(1) This paper, (2) \citet{Bryan2020}, (3) \citet{Xuan2020}, (4) \citet{Bryan2018}, (5) \citet{Snellen2014}, (6) \citet{Zhou2016}, (7) \citet{Zhou2019}, (8) \citet{Manjavacas2019}, (9) \citet{Miles-Paez2019}, (10) \citet{Zhou2020}, (11) \citet{Crossfield2014}, (12) \citet{Mohanty2005}, (13) \citet{Gagne2017}, (14) \citet{Schneider2018}, (15) \citet{Bowler2017}, (16)\citet{Kraus2014}, (17) \citet{Bowler2014}, (18) \citet{Bowler2016}, (19) \citet{Kuzuhara2011}, (20) \citet{Dupuy2018}, (21) \citet{Bowler2013}, (22) \citet{Luhman2006}, (23) \citet{Bertout2007}, (24) \citet{Currie2013}, (25) \citet{Gauza2015}, (26) \citet{Stone2016}, (27) \citet{Ireland2011}, (28) \citet{Lachapelle2015}, (29) \citet{Lagrange2010}, (30) \citet{Dupuy2019}, (31) \citet{Chauvin2005}, (32) \citet{Bell2015}, (33) \citet{Chauvin20052}, (34) \citet{Goldman2010}, (35) \citet{Scholz2010}, (36) \citet{Metchev2006}, (37) \citet{Miles-Paez2017}, (38) \citet{Bailey2014}, (39) \citet{AlvesdeOliveira2012}, (40) \citet{Gagne2015}, (41) \citet{Shkolnik2017}, (42) \citet{Lodieu2008}, (43) \citet{Allers2016}, (44) \citet{Rice2010}, (45) \citet{Kurosawa2006}, (46) \citet{Faherty2016}   }
\end{deluxetable}
\end{turnpage}
\clearpage

\noindent \citep{Schwarz2016,Zhou20182,Vos2020}, their higher masses exclude them from our sample.  

In subsequent sections we primarily consider correlations not just with rotation rate alone but with how fast each object is spinning relative to their break-up speed, $v/v_{break}$.  This quantity is particularly illustrative because in the absence of any braking mechanism, we would expect each object to have spun up to break-up speeds at young ages when they were actively accreting material and angular momentum.  This fractional break-up speed $v/v_{break}$ is therefore an estimate of the extent to which angular momentum has been extracted from each object, which in turn can shed light on physical braking mechanisms at play.  We calculate break-up speed and uncertainties using equations \ref{eq:v_b} and \ref{eq:sigma v_b}

\begin{equation}
v_{break} = \sqrt{\frac{GM}{R}}
\label{eq:v_b}
\end{equation}

\begin{equation}
\sigma_{v_{break}} = \frac{v_{break}}{2}\sqrt{\bigg(\frac{\sigma_{M}}{M}\bigg)^2 + \bigg(\frac{\sigma_{R}}{R}\bigg)^2}
\label{eq:sigma v_b}
\end{equation}

\noindent where relevant values are shown in Table \ref{tb:res}. We note that because the planet's polar and equatorial radii differ at break-up, the actual break-up velocity is $\sqrt{2/3}v_{break}$ \citep[e.g.,][]{Porter1996}.
The difference between $\sqrt{2/3} \simeq 0.82$ and 1 
is small compared to model and data uncertainties and so we drop this correction for simplicity.

\section{Results and Discussion}

\subsection{Spin evolution in the planetary-mass regime}\label{sec:spin_evolution}

\begin{figure*}
    \epsscale{1.1}
    \centering
    \plottwo{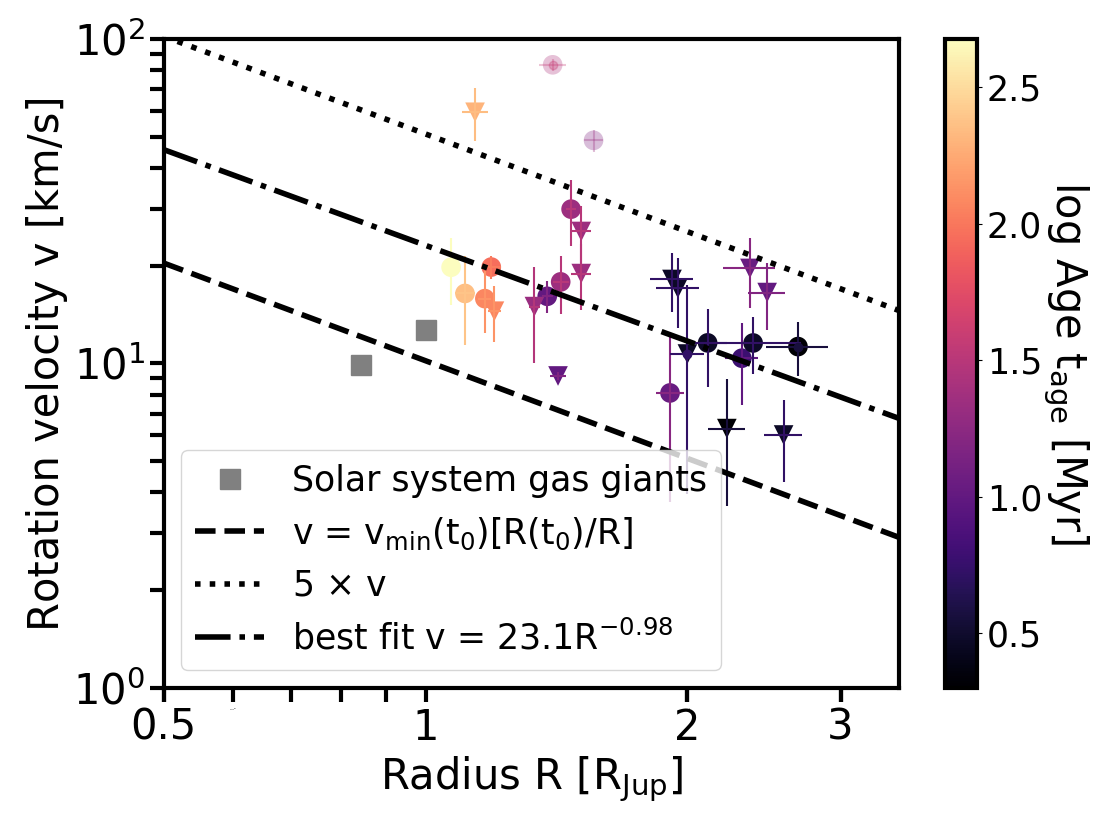}{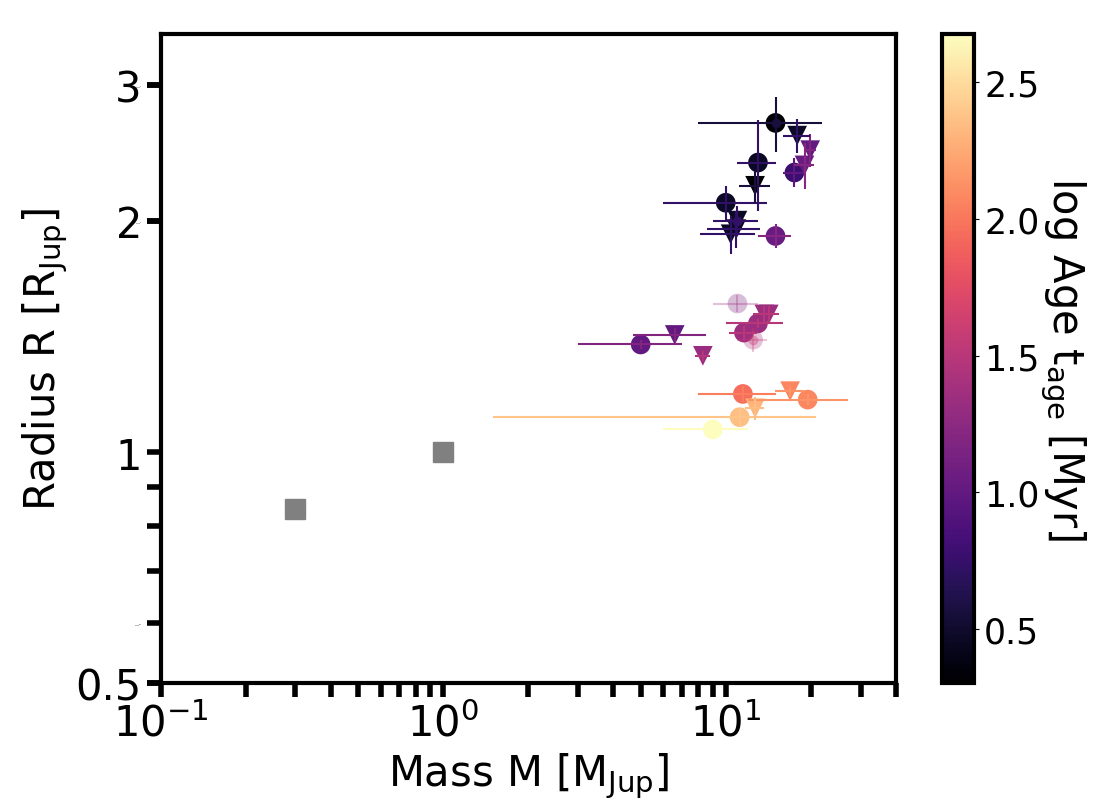} 
    \caption{Left:  Rotation velocity $v$ as a function of radius $R$.  Both bound and free-floating objects are colored according to their age $t_{\rm age}$, and the Solar System gas giants are shown as gray squares.  Isolated low-mass brown dwarfs are denoted by triangles, and bound companions are plotted as circles.  Measurements for AB Pic b and HD 106906 b are translucent because their photometric rotation periods are marginal ($<$ 3$\sigma$ significance).  We include them in Figures 5--7 for completeness but do not include these values in our interpretation.  There is an anti-correlation between $v$ and $R$, where younger objects have larger radii and rotate more slowly.  The black dashed line corresponds to $v = v_{\rm min}(t_0)(R(t_0)/R)$, where $t_0$ = 1 Myr (assumed time of disk dispersal), $v_{\rm min}(t_0)$ is the minimum value for $v$ at $t_0$ evaluated at $R=R(t_0)$ and $M = 10 M_{\rm Jup}$ using eq. \eqref{eq:v_min_t0}, and $R$ is the radius of the PMO. The dotted line is $v$ increased by a factor of five.  The dot-dash line shows the best-fit $v ({\rm km/s}) = C_1 \times (R (R_{\rm Jup}))^{C_2}$ relation, where $C_1$ = 23.1 and the slope $C_2$ = -0.98.  Right:  Radius $R$ as a function of mass $M$ (same colors and markers). The PMOs in our sample have comparable masses.
    \label{fig: radius age}}
\end{figure*}

With a sample of 27 spin measurements distributed over a variety of ages, we can begin to piece together how spin evolves in the planetary-mass regime.   Figure \ref{fig: radius age} shows that our objects, all of which have masses of approximately 10 $M_{\rm Jup}$ (right panel), appear to shrink in radius and increase their rotational velocities as they age from $\sim$2 to $\sim$400 Myr (left panel).  To quantify this trend and its significance, we fit the following relation: $v = C_1\times R^{C_2}$, which in log space yields $\log(v) = \log(C_1) + C_2\times \log(R)$. Using an MCMC framework, 
we maximize the likelihood function

\begin{equation}
\log{L} = \sum_{i=1}^{n} -0.5\bigg(\frac{(m_i - d_i)^2}{(\sigma_{i})^2 + (\sigma_{\rm jit})^2}\bigg) -0.5\log(2\pi [\sigma_i^2 + \sigma_{\rm jit}^2]),
\label{eq:loglike2}
\end{equation}

\noindent where $m_i$ is the $i$th model rotation speed $m = C_1\times R^{C_2}$, $d_i$ is the $i$th measured rotational velocity $v$, $\sigma_i$ is the measurement uncertainty on the spins, and $\sigma_{\rm jit}$ is a jitter term which accounts for potential astrophysical variance in our sample.  The model parameters $C_1$ and $C_2$ are free, as is $\sigma_{\rm jit}$ . 

We show the best-fit relation in Figure \ref{fig: radius age}, where the slope of the line is $C_2 = -0.98^{+0.75}_{-0.68}$, and the jitter term is $\sigma_{\rm jit} = 1.0^{+0.8}_{-0.4}$ km/s.   The slope of this trend is consistent with angular momentum conservation: rotational velocities increase as $v \propto 1/R$. As objects cool and contract, they spin up.

\begin{figure}[h]
\centering
\includegraphics[width=0.48\textwidth]{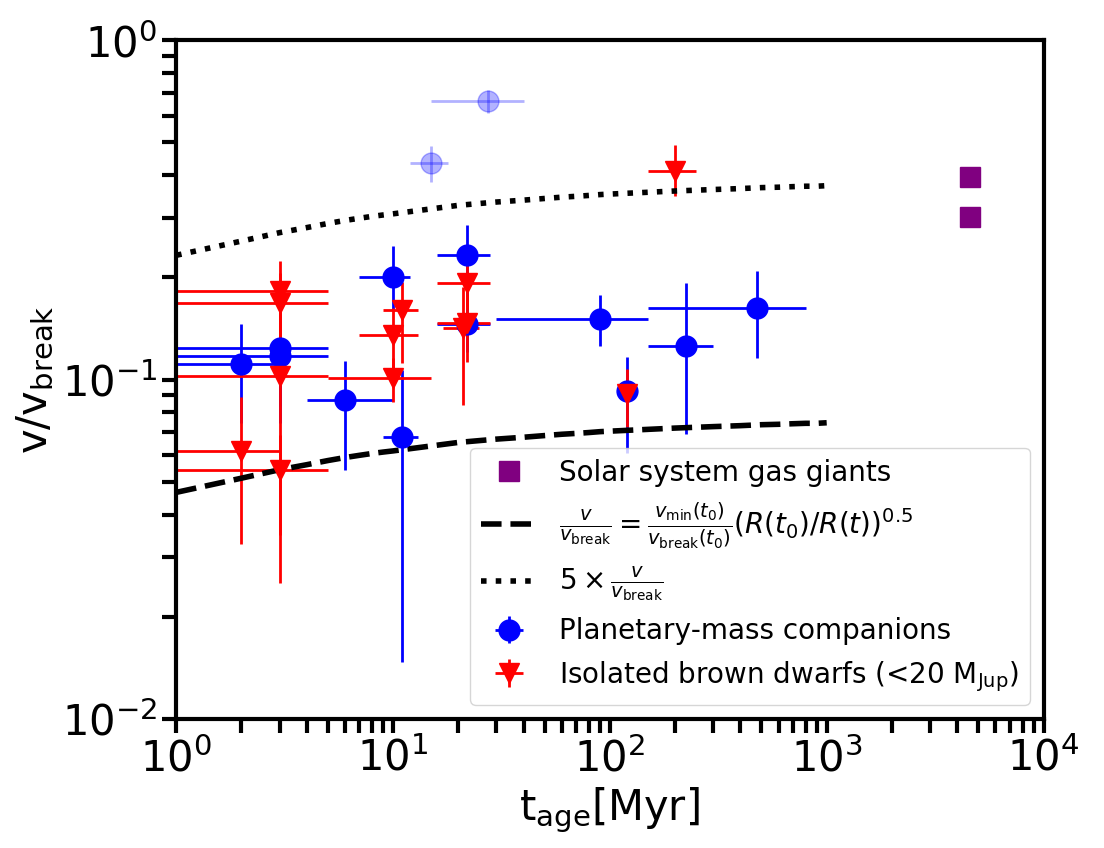} 
\caption{Rotation rates as a fraction of break-up ($v_{\rm break} \equiv \sqrt{GM/R}$), plotted for our sample of 14 planetary-mass companions (blue squares), 13 free-floating low-mass brown dwarfs (red triangles), and Jupiter and Saturn (purple squares).  We show 1$\sigma$ uncertainties for all objects, which are dominated by uncertainties on break-up speeds but also include errors from the directly measured rotation speeds as well as the unknown inclination angle relative to our line of sight.  Measurements for AB Pic b and HD 106906 b are translucent because their photometric rotation periods are marginal ($<$ 3$\sigma$ significance).  We include them in Figures 5--7 for completeness but do not include these values in our interpretation. While the two lines shown here are simply the same two in Figure \ref{fig: radius age} mapped onto these new axes, they are more shallow because both $v$ and $v_{\rm break}$ increase with increasing age ($v/v_{\rm break} \propto 1/R^{0.5}$ as opposed to $v \propto 1/R$).  Rotational velocities appear to be set early, at or before an age of a few Myr, to about a tenth of break-up. 
}
\label{fig: ang mom}
\end{figure}

What sets the normalization of this trend, i.e., what sets the angular momentum budget? To investigate this, we normalize the spins to their break-up values, and plot them vs.~age in Figure \ref{fig: ang mom}. We find that our sample of PMOs rotate about an order of magnitude slower than break up, and that this behavior is established at an early age -- the youngest objects in our sample are 2$\pm$1 Myr old.  These observational results are consistent with spin angular momentum being set during the era of planet formation, and in particular with the theory of disk locking: when a planet is still surrounded by a circum-PMO disk (CPD), magnetic torques exerted by the CPD on the planet can spin the latter down
\citep[][hereafter GC20]{Takata1996,Batygin2018,Ginzburg2020}. Without shedding angular momentum, the planet would cool and contract until it attained break-up velocity, after which it would be unable to contract further. One way to shed angular momentum is by expelling mass at the equator when break-up is reached, forming a CPD \citep{WardCanup2010}; subsequent magnetic interaction with the CPD, and transport of angular momentum from the CPD to larger distances, can then reduce the rotation significantly below break-up. GC20 found that the magnetic braking time-scale is shorter than the planet's Kelvin-Helmholtz contraction time, enabling planets to overcome the angular momentum barrier and contract toward their zero-temperature sizes (on the order of a Jupiter radius) with rotation velocities significantly below break-up, as observed. 

While bound companions and isolated low-mass brown dwarfs likely have different formation histories,
with bound objects typically forming in a disk and isolated ones forming via molecular cloud fragmentation, both routes should produce accretion disks around planetary-mass objects. In the case of a companion,
the angular momentum to form a circum-PMO disk comes from the circum-primary disk, while in the isolated case, the angular momentum originates from the turbulent molecular cloud. Disk locking theory applies to either scenario. The fact that the spin distributions between bound and isolated populations are consistent with each other \citep[e.g.,][]{Bryan2018,Xuan2020} supports a common mechanism.

In the magnetic braking theory, the planet's rotation velocity is regulated to match the Keplerian orbital velocity at the disk's magnetospheric truncation radius.  By the time the disk disperses at an age $t_0$, the planet's rotational velocity, relative to break-up, reaches a minimum value given by eq. 9 of GC20:

\begin{equation}\label{eq:v_min_t0}
\left.\frac{v_{\rm min}}{v_{\rm break}}\right\vert_{t=t_0}\sim\left(\frac{t_0\times v_{\rm break}}{2\pi R(t_0)}\right)^{-1/7}   
\end{equation}
where $v_{\rm break}$ is the break-up velocity
\begin{equation}
v_{\rm break}=\left[\frac{GM}{R(t_0)}\right]^{1/2}.    
\end{equation}
We calculate $R(t)$ using the hot-start evolutionary model grid \citep{Burrows1997}, which was also used in section 3.5.  For $t>t_0$, the planet spins up as it contracts ($R(t) < R(t_0)$) at constant angular momentum:

\begin{equation}
\begin{split}
    v = \left.v_{\rm min}\right\vert_{t=t_0} \times \left[\frac{R(t_0)}{R(t)}\right]=\\
    \left(\frac{t_0}{2\pi R(t_0)}\left[\frac{GM}{R(t_0)}\right]^{1/2}\right)^{-1/7}\left[\frac{GM}{R(t_0)}\right]^{1/2}\left[\frac{R(t_0)}{R(t)}\right].
\end{split}
\label{eq: contract}
\end{equation}

Equation (\ref{eq: contract}) is plotted as a dashed line in Figures \ref{fig: radius age} and \ref{fig: ang mom}. We see that it correctly bounds our data. GC20 treat $v_{\rm min}$ as a lower limit because they find that the magnetic braking time-scale is shorter than the cooling time by only a small margin. Order-unity uncertainties, due, e.g., to the complex transition from a disk geometry to a more spherical magnetosphere at the truncation radius, could reduce the braking efficiency and prevent the planet from spinning down all the way to the asymptotic $v_{\rm min}$ (which was calculated in GC20 by omitting order-unity coefficients). These inefficiencies, which also include the degree to which these CPDs are ionized and magnetically coupled to planets, and how much angular momentum CPDs can transfer to circumstellar material at large \citep{Batygin2018}, could account for why rotational velocities plot consistently above the $v_{\rm min}$ line, and why they exhibit a factor of
4--5 scatter. Another contribution to the scatter could include 
variations in CPD lifetime -- the dependence of velocity on time of disk dispersal $t_0$ is evident in equation \ref{eq: contract}, where a larger $t_0$ decreases the velocity both directly ($v\propto t_0^{-1/7}$) and indirectly through $R(t_0)$ ($v\propto R(t_0)^{5/7}$, where $R(t_0)$ is smaller for larger $t_0$).  Assuming outside values for $t_0$ of 1 Myr and 10 Myr, and calculating
$R(t_0)$ at fixed $M = 10 M_{\rm Jup}$ using the model of \citet{Burrows1997}, we find that $v/v_{\rm break}$ varies by a factor of $\sim$2. 

Figure \ref{fig: ang mom} argues that as PMOs age after CPD dispersal, they do not lose angular momentum.  We see no signature of, e.g.,  spin-down due to magnetized winds, which play a substantial role in extracting angular momentum from stars and to a lesser extent brown dwarfs \citep{Gallet2013,Gallet2015,Zapatero2006,Scholz2015}. However, with the exception of the Solar System planets Jupiter and Saturn,  our data do not extend for ages longer than $\sim$400 Myr.

As for the angular momentum evolution of low-mass stars and brown dwarfs at earlier ages, spins in this higher-mass regime extend
from $\sim$10\% of break-up all the way
up to break-up \citep{Moore2019}. This factor of 10 scatter is set $\lesssim$a few Myr for these higher-mass objects, comparable to our findings in the planetary-mass regime.  There is evidence that disk locking plays a critical role in regulating the angular momenta of low-mass stars and brown dwarfs --- the median rotational period for objects observed to still host disks is at least 50$\%$ longer than those objects whose disks have  dispersed \citep{Moore2019,Scholz2018}.  

\subsection{No correlations with $a$ or M/M$_{\rm star}$}

\begin{figure}[h]
\begin{tabular}{ c }
\includegraphics[width=0.48\textwidth]{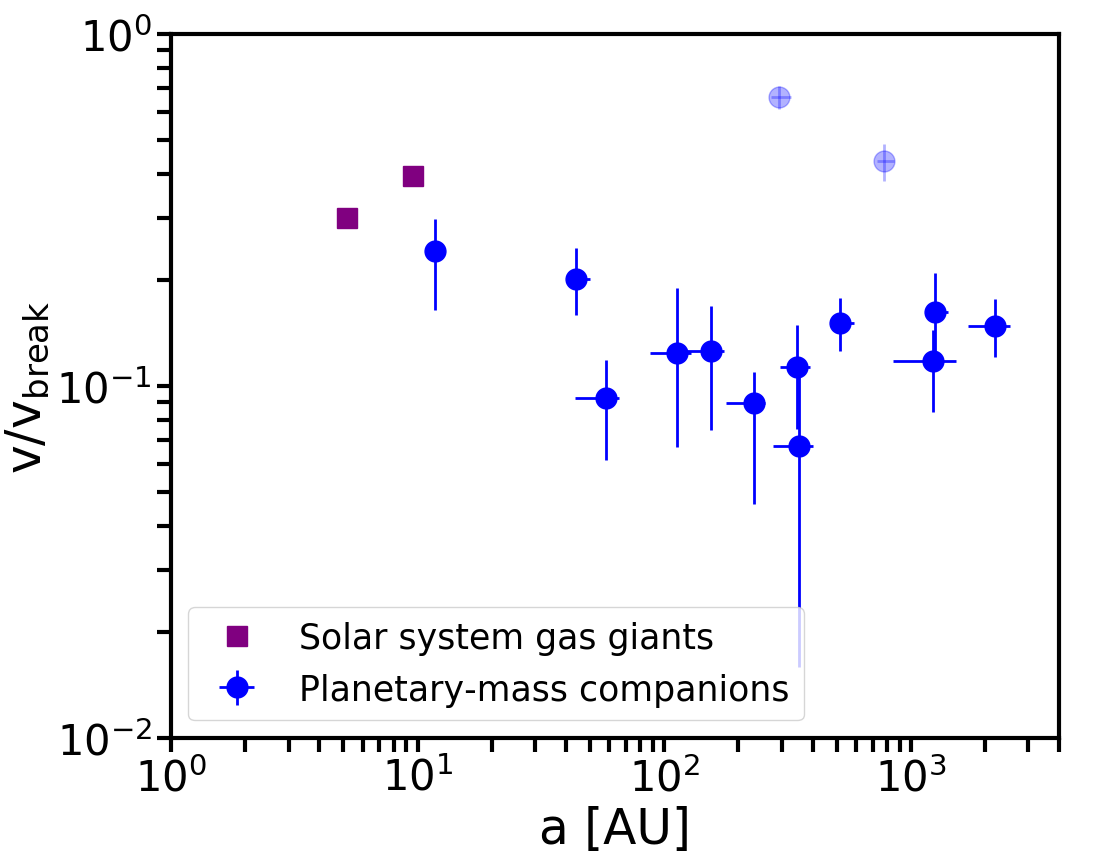} \\
\includegraphics[width=0.48\textwidth]{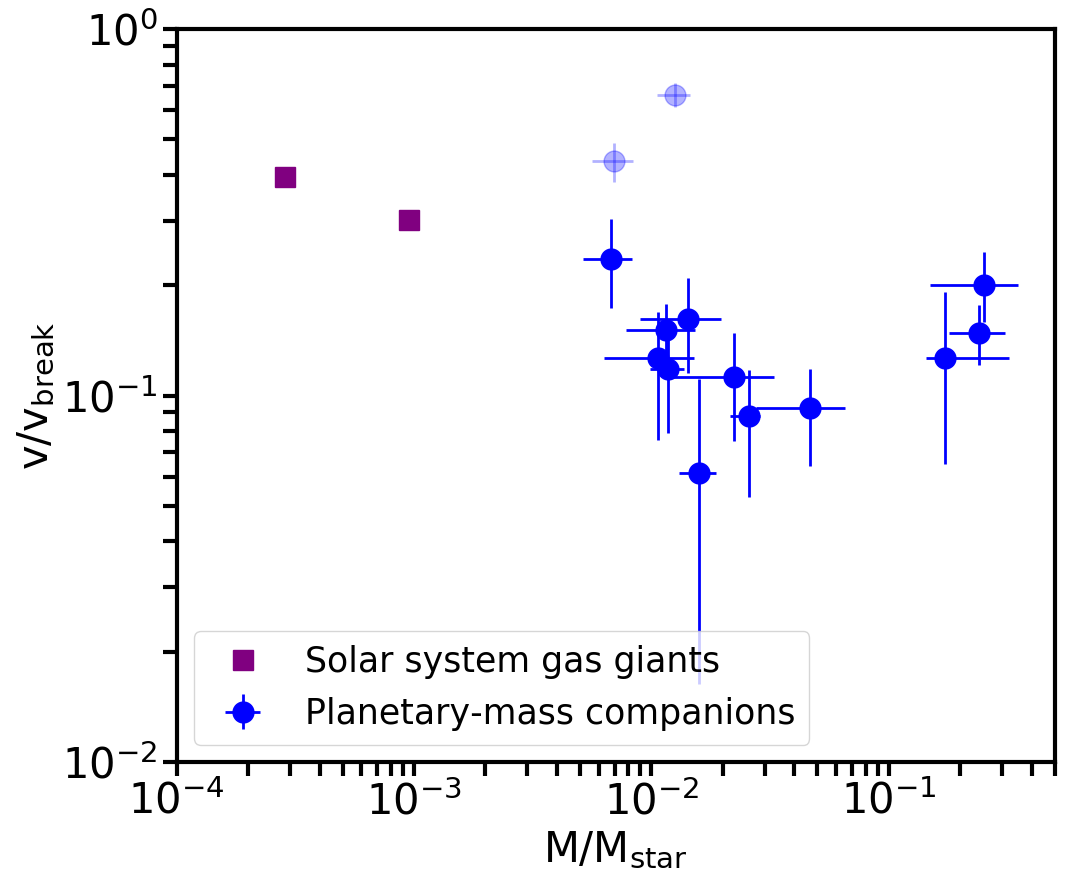}
\end{tabular}
\caption{Top panel:  Fractional break-up velocity as a function of semi-major axis.  Planetary-mass companions are shown in blue and Jupiter and Saturn are denoted by purple squares. Measurements for AB Pic b and HD 106906 b are translucent because their photometric rotation periods are marginal ($<$ 3$\sigma$ significance).  There is no significant correlation between $v/v_{\rm break}$ and PMC separation from their host stars.  This is unsurprising given that the bolometric luminosities of these companions are orders of magnitude higher than the power they receive from their host stars; energy budgets and cooling histories are
set locally. Bottom panel:  Rotation rate as a fraction of break-up velocity is plotted as a function of system mass ratio.  There is no significant correlation between $v/v_{\rm break}$ and $M/M_{\rm star}$ either with or without the solar system gas giants.  This is unsurprising given that the disk locking mechanism operates largely independent of formation scenario.}
\label{fig: sema}
\end{figure}

In Figure \ref{fig: sema} we show $v/v_{\rm break}$ as functions of semi-major axis $a$ (top panel) and mass ratio $M/M_{\rm star}$ (bottom panel) for the bound PMOs in our sample. Neither data set shows a significant correlation.

The lack of a trend is consistent with disk locking theory and subsequent contraction at constant angular momentum, as described in the previous subsection.  Angular momentum budgets are determined locally by CPDs, which are heated and ionized by their host PMO luminosities, not by the central stars whose radiative fluxes are negligible for the wide orbits considered here.

The lack of a correlation is also consistent with the idea that spin regulation is independent of  formation scenario. Core accretion is more efficient at close separations, and is typically believed to form planets with small $M/M_{\rm star}$, whereas gravitational instability is believed to be the dominant formation channel far from the star, producing companions with larger mass ratios \citep[e.g.,][]{Nielsen2019,Bryan2020}. The fact that bound PMO spins do not correlate with either $a$ or $M/M_{\rm star}$ --- and are similar even to isolated PMO spins --- is consistent with disk locking theory, which can operate regardless of whether the PMO is bound or isolated, and regardless of whether objects form from the bottom-up or top-down, as all formation scenarios involve a disk orbiting the PMO.  GC20 showed that PMO spins are determined by magnetic interaction with the innermost edges of disks at late times, i.e., for times near the disk dispersal time $t_0$. This late-time interaction is insensitive to earlier initial conditions or to environmental conditions far from the PMOs.

\subsection{Presence of CPDs?}
The presence of a CPD likely plays a central role in the spin evolution of PMOs -- as long as an object hosts a CPD, angular momentum can be extracted from the system.  A relevant question is therefore:  Do any of our PMOs, in particular the youngest ($\lesssim$ 10 Myr), still host disks?
In our sample of 14 bound companions, 6 have ages $\lesssim 10$ Myr:  ROXs 12 b, ROXs 42B b, DH Tau b, GSC 6214-210 b, SR 12 c, and 2M1207 b.  Near-infrared imaging of ROXs 12 b showed anomalously red K'-L', which might be explained by the presence of a disk \citep{Kraus2014}. There have been no disk markers observed for ROXs 42B b, both from 3-5 $\mu$m photometry which did not show excess thermal emission that could point to a surrounding disk \citep{Daemgen2017}, and from JHK spectra which did not exhibit accretion indicators like Pa$\beta$ emission \citep{Bowler2014}. A NIR spectrum of SR 12 c showed no evidence of accretion indicators in JHK \citep{Bowler2014}. On the other hand, multiple observations of DH Tau b strongly suggest that it continues to host a CPD.  In addition to having significant H$\alpha$ and Pa$\beta$ emission in the optical and NIR, there is an optical continuum excess \citep{Bonnefoy2014,Zhou2014}.  NIR and optical spectra of GSC 6214-210 b reveal multiple accretion indicators, namely Pa$\beta$ emission, Br$\gamma$ emission, and H$\alpha$ emission, as well as an optical continuum excess \citep{Bowler2014,Zhou2014,Lachapelle2015}.    Finally, no observations have found evidence of a disk around 2M1207-3932 b \citep{Skemer2014,Ricci2017}.  

These indirect indicators of active accretion and/or the presence of IR/optical continuum excess have motivated efforts to directly detect these disks using ALMA.  Of the six PMCs in our sample $\lesssim$ 10 Myr old, five have prior ALMA observations:  ROXs 42B b \citep{Wu2020}, ROXs 12 b \citep{Wu2020}, DH Tau b \citep{Wu2017,Wu2020}, GSC 6214-210 b \citep{Bowler2015,Wu2017}, and 2M1207-3932 b \citep{Ricci2017}. None of these observations detected disks around these objects.  More generally,  multiple programs targeting 19 PMCs \citep[see][and references therein]{Perez2019, Wu2020} have yet to unambiguously detect a CPD, with the possible exception of PDS 70 c \citep{Isella2019}.  Perhaps CPDs have evaded detection in the millimeter because they are optically thin, or because they are more compact than originally envisioned (see, e.g., \citealt{Zhu2016}; or \citealt{Fung2019} who find that CPDs are rather small, extending to only $\sim$10--20$\%$ of the Bondi radius when the Bondi radius is smaller than the Hill radius).

Among the 13 isolated PMOs, 8 have ages $\lesssim$10 Myr:  OPH 98, OPH 103, OPH 90, USco 1608-2315, 2M1207-3932, GY 141, KPNO Tau 12, and WISE 1147-2040.  All three members of $\rho$ OPH (98, 103, 90) show mid-IR excesses, indirectly indicating the presence of a disk \citep{AlvesdeOliveira2012}. In addition, H$\alpha$ emission from KPNO 12 and 2M1207-3932A indicate active accretion \citep{Mohanty2005}, and ALMA observations directly detect a compact disk around 2M1207-3932A \citep{Ricci2017}. In contrast, USco 1608-2315 \citep{Lodieu2008}, WISE 1147-2040 \citep{Faherty2016}, and GY 141 \citep{Mohanty2005} show no evidence for a disk.  Out of these 8 objects, only 2M1207-3932A has been targeted with ALMA to directly detect the presence of CPDs.

To summarize, 2 of the 6 youngest PMCs in our sample show evidence of residual CPDs, as do 5 of the 8 youngest free-floating brown dwarfs; the rest do not, but this may be because their CPDs are too faint. A larger sample of PMOs with CPDs would enable us to study how disk locking works in ``real time'', and might shed light on whether disk dispersal timescales could significantly impact resulting rotational velocities.

\section{Conclusions}

In this study we measure rotational line broadening using near-IR high-resolution spectra from NIRSPEC/Keck for eight planetary-mass objects (PMOs). These measurements are combined with previously published rotational velocities to create a catalog of 27 spin measurements, of which 14 are for bound planetary-mass companions and 13 for free-floating low-mass brown dwarfs.  Sixteen of these measurements are from our NIRSPEC/Keck program \citep[][this paper]{Bryan2018,Xuan2020,Bryan2020}.  

We find that as objects age and cool and their radii $R$ contract, their rotational velocities scale as $v \propto 1/R$. Thus we find that spin angular momentum is conserved over the ages spanned by our sample, from $\sim$2 to $\sim$400 Myr (5 Gyr if we include Jupiter and Saturn). Angular momentum budgets appear set at earlier ages, and are such that objects rotate below break-up speeds by about an order of magnitude.  More specifically, at ages $\lesssim 10$ Myr, PMOs spin at rates 5--20\% of  break-up, consistent with their rotation having been regulated by magnetic torques exerted by circum-PMO disks (CPDs). We speculate that such CPDs could be the breeding grounds for satellite systems like the Galilean moons \citep{CanupWard2002,CanupWard2006,BatyginMorbidelli2020}.

Future directions include measuring spins for younger objects that still host disks.  Inferring disk and PMO rotation periods using light curves
would enable us to test the disk locking hypothesis, as has been done for pre-main-sequence stars (e.g., Stauffer et al. 2015; Ansdell et al. 2016; Bodman et al. 2016).  We should also increase our sample
size at the oldest ages. For stars and to a lesser extent higher-mass brown dwarfs, spin down due to magnetized winds is seen after a few hundred Myrs,
and it would be interesting to see if we can observe a similar effect for
PMOs; compare Fig. 7 in \citet{Gallet2013} to our Fig. \ref{fig: ang mom}. Finally, our sample of PMOs has a fairly narrow range of masses around $10 M_{\rm Jup}$; pushing down in mass would allow us to explore how spin regulation depends on mass.

\section*{}
We thank Ian Czekala, Daniel Fabrycky, and Yifan Zhou for helpful conversations.  M.L.B. and S.G. are supported by their Heising-Simons Foundation 51 Pegasi b Fellowships. This work benefited from NASA’s Nexus for Exoplanet System Science (NExSS) research coordination network sponsored by the NASA Science Mission Directorate, and was supported in part by NASA grant NNX15AD95G/NEXSS.  We extend special thanks to those of Hawaiian ancestry on whose sacred mountain of Mauna Kea we are privileged to be guests.

\bibliographystyle{aasjournal}
\bibliography{bibliography}

\end{document}